\documentclass[aip,graphicx]{revtex4}

\usepackage{graphicx}  
\usepackage{subfigure}
\usepackage{multirow}
\usepackage{xcolor}
\usepackage{fancyhdr}
\usepackage{longtable}
\usepackage{parskip}
\usepackage[T1]{fontenc}
\usepackage{dcolumn}   

\usepackage{bm}        
\usepackage{amsfonts}  
\usepackage{amsmath}   
\usepackage{amssymb}   
\usepackage{soul}
\usepackage[normalem]{ulem}

\newcommand{\pwisein}{\left\{ \begin{array}{ll}}
\newcommand{\pwiseout}{\end{array}\right.}

\let\ref\cref
\usepackage{hyperref} 
\usepackage{cleveref}

\begin{document}
\title{Diffusion dynamics of an overdamped active ellipsoidal Brownian particle in two dimensions}

\author{Sudipta Mandal}
\email{sudiptomandal94@gmail.com}
\affiliation {Department of Physical Science, Indian Institute of Science Education and Research Mohali,Sector 81, S. A. S Nagar,Manauli PO 140306,India}
\author{Anirban Ghosh}
\email{anirbansonapur@gmail.com}
\affiliation {Raman Research Institute, Bengaluru, 560080, India}
\affiliation{Department of Physical Science, Indian Institute of Science Education and Research Mohali,Sector 81, S. A. S Nagar,Manauli PO 140306,India}

\date{\today}

\begin{abstract}  

Shape asymmetry is the most abundant in nature and attracted great interest in recent research. The phenomenon is widely recognized: a free ellipsoidal Brownian particle displays anisotropic diffusion during short time intervals, which subsequently transitions to an isotropic diffusion pattern over longer time scales. We have further expanded this concept to incorporate active ellipsoidal particles characterized by an initial self-propelled velocity. This paper provides analytical and simulation results of diffusion dynamics of an active ellipsoidal particle. The active ellipsoidal particle manifests three distinct regimes in its diffusion dynamics over time. In the transient regime, it displays diffusive behavior followed by a super-diffusive phase, and in the longer time duration, it transitions to purely diffusive dynamics. We investigated diffusion dynamics for the free particle as well as the particle in a harmonic trap, and the particle subject to a constant field force. Moreover, we have studied the rotational diffusion dynamics and torque production resulting from an external constant force field. Furthermore, our investigation extends to the examination of the scaled average velocity of an ellipsoidal active particle, considering both a constant force field and a one-dimensional ratchet. 
 
\end{abstract}

\maketitle 

\section{Introduction}


Self-propulsion stands as a significant accomplishment of biological evolution, playing a crucial role in the survival of various living species, including bacteria, cells, algae, and other micro-organisms. This ability enables these organisms to engage in essential activities such as locomotion in search of nourishment, orientation towards light, and the dissemination of their own species. At the mesoscopic scale, the mechanism behind self-propulsion has long been associated with either the infusion of external energy or the internal conversion of chemical energy. As a result of this non-equilibrium process, fundamental equilibrium properties such as detailed balance, zero net energy flux, and the fluctuation-dissipation theorem are unable to maintain their validity. A wide array of instances showcasing self-propelled systems in nature exists, often referred to as "Active matter". These examples span across various scales, ranging from individual entities like motile cells, spermatozoa, and swimming bacteria, to more complex collective phenomena like the swarming of bees, the schooling of fishes, and the flocking of birds, among others.

The investigation of anisotropic particles has a rich history, driven by the intrinsic connection between particle shape and experimentally measurable properties\cite{kurzthaler2016intermediate, ghosh2020persistence, chakrabarty2014brownian, liao2018transport, aurell2016diffusion, fan2017ellipsoidal, yuan2022diffusion, guell2010anisotropic, roh2015analysis}. Shape-asymmetric particles are prevalent in nature, varying in size from a few nanometers to a few micrometers. This problem has garnered renewed interest due to the significant influence that shape exerts on the mechanical and electrical properties of nano-sized objects. In the past decade, spurred by advancements in particle chemistry, there has been a surge in the development of particles with enhanced transport properties, aiming to replicate natural counterparts. These synthetically engineered colloids, endowed with multifunctional properties, frequently find diverse applications in photonics, nano- and biotechnology, drug delivery, and various biomedical uses. Analyzing the Brownian motion of asymmetrical particles is notably more intricate when contrasted with the spherical case. This complexity arises from the interplay between rotational and translational motion. Specifically, the correlation between the current orientation of the particle and the instantaneous translational diffusion coefficient gives rise to anisotropic motion, particularly evident during short time intervals\cite{grima2007brownian, ghosh2020persistence}. The evolution in dynamic behavior over time occurs because rotational diffusion gradually erases the initial anisotropic translational motion of the particle. The translational diffusion coefficient at long times is equivalent to the average of the diffusion coefficients measured along the semi-axes of the ellipsoidal or cylindrical particle.

 Active systems serve as more than just an experimental testing ground; they also offer a fertile domain for exploring theories within the realm of non-equilibrium statistical physics\cite{ramaswamy2010mechanics,marchetti2013hydrodynamics, ten2011brownian, sandoval2013anisotropic} and also underpin the natural processes of life\cite{needleman2017active}. There are numerous realizations of active particles\cite{lauga2009hydrodynamics,toner2005hydrodynamics, sandoval2014effective,santra2022universal,santra2021active,li2017two,ao2014active,katyal2020coarsening,biswas2020first,solon2015active,shee2020active,howse2007self,PhysRevE.100.062116} in nature ranging from bacteria\cite{leptos2009dynamics,hill2007hydrodynamic,diluzio2005escherichia} and spermatozoa\cite{riedel2005self} to artificial colloidal micro-swimmers. Self-propulsion stands as a fundamental characteristic of the majority of living systems, enabling them to sustain metabolism and engage in motion\cite{ai2014transport}. The dynamics of self-propelled particles navigating through potentials can manifest distinct and peculiar behaviors\cite{schweitzer1998complex,tailleur2008statistical,ai2014entropic,kaiser2013capturing,fily2012athermal,bickel2013flow,mishra2012collective,czirok1999collective,ai2017transport}.  The diffusion of active particles within a cubic polymer network is described by a trapped-and-hopping mechanism, resembling the diffusion of active particles in a periodic confining potential in a recent work by Kim et al.\cite{kim2022active}. The rectification of self-propelled particles under asymmetric external potentials has garnered a lot of interest recently. Recent studies on the transport of Janus particles within periodically segmented channels have revealed that rectification effects can exhibit a significantly higher magnitude compared to those observed in conventional thermal potential ratchets\cite{PhysRevLett.110.268301,PhysRevE.87.042124,angelani2011active}. Even within symmetric potentials, directed transport can be prompted through the spatially controlled self-propulsion of particles, combined with a deliberate phase shift relative to the potential landscape\cite{PhysRevE.87.042124}. Angelani et. all made a notable discovery: when analyzing run-and-tumble particles within periodic potentials, the presence of an asymmetric potential leads to the emergence of a net drift speed\cite{angelani2011active}. All of these investigations on the active ratchet considered the active particle as the point-spherical particle.


In this paper, we present a simulation model of an active ellipsoidal particle in two dimensions. We have presented the temporal evolution of mean square displacement and diffusion coefficient to characterize the dynamics of the system. In the Sec.\ref{Theory} we present the model and we review the derivation of the Langevin equations for an active ellipsoidal rigid Brownian particle in the presence of a potential field in two dimensions. In Section \ref{result}, we present and analyze the variations in mean square displacement and diffusion coefficient of an asymmetrical active particle in diverse environments, encompassing a free system, a two-dimensional harmonic potential, and a constant force field. Moreover, we delve into the rotational dynamics and the corresponding torque production. Additionally, our examination extends to the investigation of scaled average velocity, considering both a constant force field and a one-dimensional ratchet potential. Finally, in Section \ref{Conclusion}, we furnish a comprehensive summary of the paper to conclude our findings.

\section{Basic Model and Methods}\label{Theory}
\begin{figure}[h]
	    \centering
	     \includegraphics[width=0.9\textwidth]{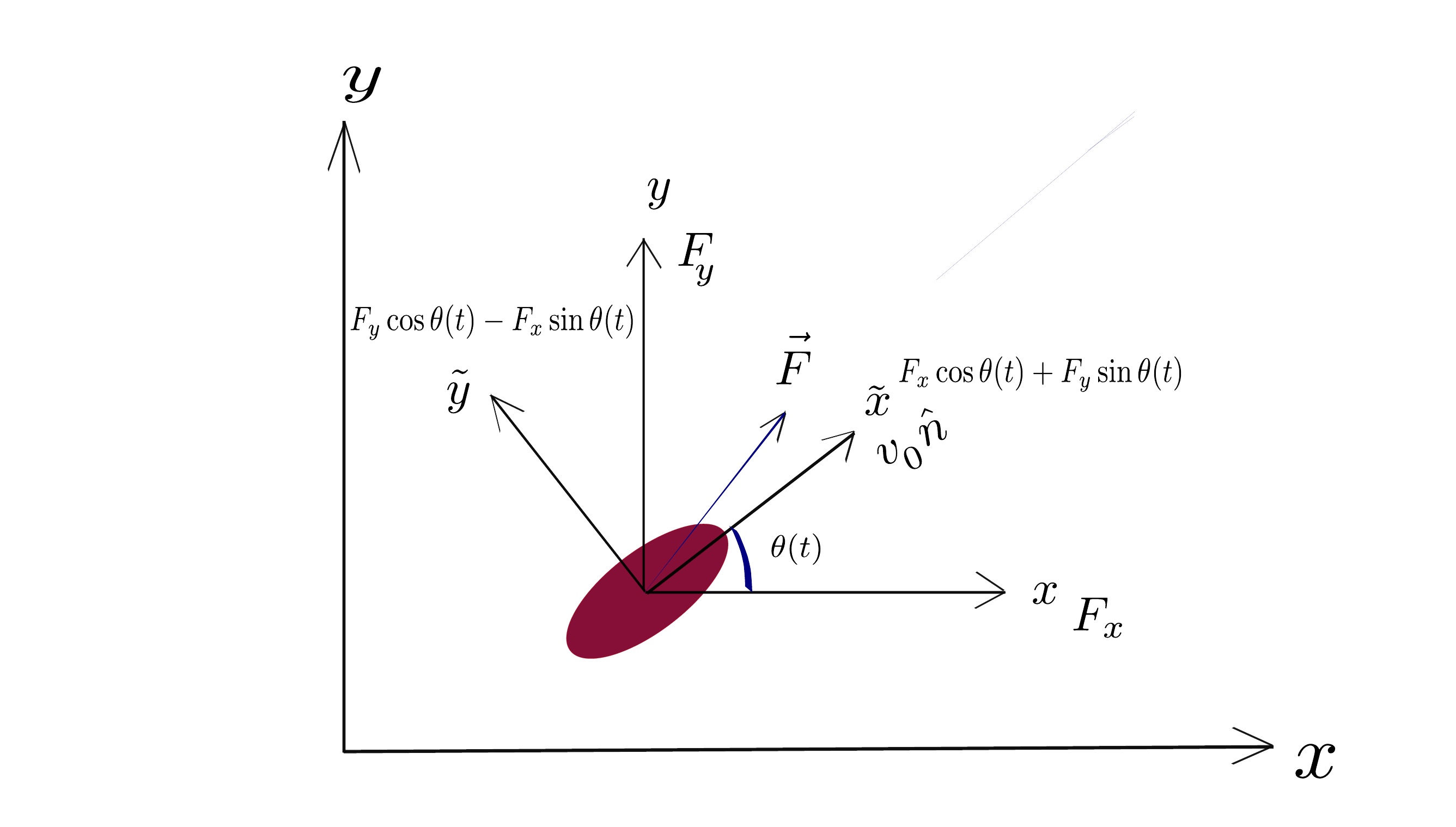}
	     \caption{ Representation of an ellipsoid in the $x-y$ lab frame and the $\tilde{x}-\tilde{y}$ body frame. The angle between two frames is $\theta$. The projections of an external force $\Vec{F}$ are shown along the $x-y$ and $\Tilde{x}-\Tilde{y}$ axes. Propulsion velocity $v_0\hat{n}$ is acting along the major axis of the particle as shown in the figure.}
	     \label{ellipse}
	    \end{figure}

We have examined the motion of an ellipsoidal particle within a potential field as shown in fig.(\ref{ellipse}). Our analysis has been confined to a two-dimensional framework. Here the particle has propulsion velocity along the longer ($\Tilde{x}$) axis. In the two-dimensional context, the orientation of the particle assumes significance and is described by a unit vector $\hat{n}_i = (\cos{\theta}_i, \sin{\theta}_i)$ in the $x-y$ plane, defining the direction of the propulsion velocity. In the lab $x-y$ frame, the particle's state at a specific time $t$ can be characterized by the position vector $R(t)$ of its center of mass. This vector also corresponds to the coordinates within the body frame ($\delta\Tilde{x},\delta\Tilde{y}$). The angle $\theta(t)$ represents the orientation between the $x$ axis of the laboratory frame and the $\Tilde{x}$ axis of the body frame. In the case of an ellipsoid particle, the frame of reference plays an important role. In the body frame, rotational and translational motion in the body frame are always de-coupled. The dynamics of the active ellipsoid particle are described by the Langevin equation shown by Bao-Quan Ai et al.\cite{ai2014transport}

\begin{equation}
\begin{split}
&\frac{1}{\Gamma_{\parallel}}\frac{\partial\Tilde{x}}{\partial t}=F_x\cos\theta(t)+F_y\sin\theta(t)+v_0/\Gamma_{\parallel}+\tilde{\eta}_1(t)\\
&\frac{1}{\Gamma_{\perp}}\frac{\partial\Tilde{y}}{\partial t}=F_y\cos\theta(t)-F_x\sin\theta(t)+\tilde{\eta}_2(t)\\
&\frac{1}{\Gamma_{\theta}}\frac{\partial\theta(t)}{\partial t}=\tau(t)+\tilde{\eta}_3(t)
\end{split}
\label{eqn:ellip3}
\end{equation}

where $v_0$ is the propulsion velocity in the body frame, which is taken along the longer axis of the ellipsoidal particle. Here $\Gamma_{\parallel}=\frac{D_{\parallel}}{k_BT}$ and $\Gamma_{\perp}=\frac{D_{\perp}}{k_BT}$ are the mobilities along its longer and shorter axes respectively. Here $\Gamma_{\theta}=\frac{D_\theta}{k_BT}$ is the rotational mobility and $\tau$ is the torque acting on the body due to its orientation relative to the direction of the potential. Here $D_{\parallel}, D_{\perp}$ and $D_{\theta}$ are the diffusion coefficients along parallel, perpendicular, and rotational axes respectively. $F_x$ and $F_y$ are the projections of an external force $\Vec{F}$ along the x and y directions of the lab frame respectively. The correlations of the thermal fluctuations in the body frame are described as

\begin{equation}
\begin{split}
&\langle \tilde{\eta}\rangle=0\\
&\langle \tilde{\eta}_i{(t)}\tilde{\eta}_j(t^{\prime})\rangle =2k_BT\Gamma_i\delta_{ij}\delta(t-t^{\prime})
\end{split}
\end{equation}

 We proceed to derive these equations in the lab frame by performing a simple coordinate rotation. The displacement in the two frames is connected through the following equations:

\begin{equation}
\begin{split}
&\delta x= \cos\theta\delta\Tilde{x}-\sin\theta\delta\Tilde{y}\\
&\delta y= \sin\theta\delta\Tilde{x}+\cos\theta\delta\Tilde{y}
\end{split}
\end{equation}

Dividing the above equations by $\delta t$, taking the limit $\delta t\rightarrow 0$,
and substituting the linear and angular velocities in the body
frame from Eq.(\ref{eqn:ellip3}), we get the final equations describing Brownian motion in the lab frame as follows:

\begin{equation}
\begin{split}
&\frac{\partial{x}}{\partial t}=v_0\cos\theta(t)+F_x[\bar{\Gamma}+\frac{\Delta\Gamma}{2}\cos2\theta(t)]+\frac{\Delta\Gamma}{2} F_y\sin2\theta(t)+\eta_1(t)\\
&\frac{\partial{y}}{\partial t}=v_0\sin\theta(t)+F_y[\bar{\Gamma}-\frac{\Delta\Gamma}{2}\cos2\theta(t)]+\frac{\Delta\Gamma}{2} F_x\sin2\theta(t)+\eta_2(t)\\
&\frac{\partial{\theta}}{\partial t}=\Gamma_{\theta} \tau(t)+\eta_3(t)
\end{split}
\label{lab_frame}
\end{equation}
The quantities $\bar{\Gamma}=\frac{1}{2}(\Gamma_{\parallel}+\Gamma_{\perp})$ and $\Delta\Gamma=(\Gamma_{\parallel}-\Gamma_{\perp})$
denote the average and the difference between the mobilities of the body, respectively. The parameter $\Delta\Gamma$ represents the asymmetry of the body, for the particle of perfect spherical nature $\Delta\Gamma = 0$. In the component form, the mobility tensor is given by

\begin{equation}
\Gamma_{ij}=\bar{\Gamma}\delta_{ij}+\frac{\Delta\Gamma}{2}
\begin{pmatrix}
\cos2\theta & \sin2\theta \\ \sin2\theta & -\cos2\theta
\end{pmatrix}
\end{equation}
where $\Bar{\Bar{\mathcal{R}}}$, the correlation matrix is represented as 
\begin{equation*}
\Bar{\Bar{\mathcal{R}}}=\begin{pmatrix}
\cos2\theta & \sin2\theta \\ \sin2\theta & -\cos2\theta
\end{pmatrix}
\end{equation*}
 Note also that movement in the $x$ and $y$ directions are not independent of each other but rather are coupled through the angular position of the particle. This effectively couples the particle$'$s translational diffusion to its rotational diffusion and the strength of this coupling behavior increases proportionally with particle shape asymmetry being zero for a spherical particle.
 
\begin{equation}
\begin{split}
&\langle\eta_3(t)\eta_3(t^{\prime})\rangle=2 D_\theta\delta(t-t^{\prime})\\
&\langle \eta_i(t)\eta_j(t^{\prime})\rangle_{\theta(t)} ^{\eta_1,\eta_2}=2 k_BT\Gamma_{ij}\delta(t-t^{\prime})
\end{split}
\label{fluctuation}
\end{equation}

It is possible to directly compute the ensemble average of $\cos \theta(t)$ and $\sin{\theta}(t)$ over the thermal variations in the degrees of freedom for direction by noting the fact $\Delta\theta=\theta(t)-\theta_0$ is a Gaussian random variable, and consequently holds the identity,
\begin{equation}
\langle e^{\pm i m\Delta\theta(t^\prime)}\rangle_{\eta_3}=e^{-m^2D_\theta t^\prime}
\end{equation}

The rotational dynamics of the ellipsoidal ABP can be modelled by the Langevin equation of Eq.(\ref{lab_frame}). The torque is generated through the application of an external force. In the case of a free particle, the rotational dynamics exhibit purely diffusive characteristics, given that the angular displacement $\theta(t)$ is subject to Gaussian stochastic noise originating from fluctuations. In the modeling of torque for an ellipsoidal ABP, the torque is expressed as 
\begin{equation}
\Vec{\tau}(t)=\Vec{r}(t)\times\Vec{F}(t)
\label{torque}
\end{equation}
 where $\Vec{F}$ represents the external force applied on the particle. The $x$ and $y$ components of this force are denoted as $F_x$ and $F_y$, respectively. Therefore, the magnitude of the torque is represented by $\tau=xF_y-yF_x$.

The model studied by Basu et al.\cite{PhysRevE.98.062121}, considered an active Brownian particle without any thermal fluctuation and an exclusive shape asymmetry. The dynamics is translated to that of an exponentially decaying two-time correlation function for a Brownian particle whose effective noise is constrained by the propulsion velocity. 

 \section{Result and Discussion}\label{result}
\subsection{Free system}

The expression of the second moments for both the $x$ and $y$ axes for an active ellipsoidal particle by solving Langevin equation defined in Eq.(\ref{lab_frame})\cite{ghosh2022persistence}
 
\begin{equation}
\begin{split}
\langle x^2(t)\rangle&=2k_BT\Bigg[\bar{\Gamma}t+\frac{\Delta \Gamma}{2}\cos 2\theta_0\Big(\frac{1-e^{-4D_\theta t}}{4D_\theta}\Big)\Bigg]+\frac{v_0^2\cos 2\theta_0}{12D_\theta^2}\big(3-4e^{-D_\theta t}+e^{-4D_\theta t}\big)+\frac{v_0^2}{D_\theta^2}\big(D_\theta t+e^{-D_\theta t}-1\big)\\
&=\Bigg[2\bar{D}t+\Delta D\cos 2\theta_0\Big(\frac{1-e^{-4D_\theta t}}{4D_\theta}\Big)\Bigg]+\frac{v_0^2\cos 2\theta_0}{12D_\theta^2}\big(3-4e^{-D_\theta t}+e^{-4D_\theta t}\big)+\frac{v_0^2}{D_\theta^2}\big(D_\theta t+e^{-D_\theta t}-1\big)
\end{split}
\label{eqn:msd1}
\end{equation}
and
\begin{equation}
\begin{split}
\langle y^2(t)\rangle&=2k_BT\Bigg[\bar{\Gamma}t-\frac{\Delta \Gamma}{2}\cos 2\theta_0\Big(\frac{1-e^{-4D_\theta t}}{4D_\theta}\Big)\Bigg]-\frac{v_0^2\cos 2\theta_0}{12D_\theta^2}\big(3-4e^{-D_\theta t}+e^{-4D_\theta t}\big)+\frac{v_0^2}{D_\theta^2}\big(D_\theta t+e^{-D_\theta t}-1\big)\\
&=\Bigg[2\bar{D}t-\Delta D\cos 2\theta_0\Big(\frac{1-e^{-4D_\theta t}}{4D_\theta}\Big)\Bigg]-\frac{v_0^2\cos 2\theta_0}{12D_\theta^2}\big(3-4e^{-D_\theta t}+e^{-4D_\theta t}\big)+\frac{v_0^2}{D_\theta^2}\big(D_\theta t+e^{-D_\theta t}-1\big)
\end{split}
\label{eqn:msd2}
\end{equation}
Here $\theta_0$ is defined as the initial orientational angle of the particle with the $x$ axis. The average displacements along the $x$ and $y$ directions are calculated as follows
 \begin{equation}
\begin{split}
\langle x(t)\rangle&=\int_{0}^{t}v_0\langle\cos{\theta(t^\prime)}\rangle dt^\prime+\int_{0}^{t}\langle\eta(t^\prime)\rangle dt^\prime\\
&=\frac{v_0\cos{\theta_0}}{D_\theta}(1-e^{-D_\theta t})\\
\langle y(t)\rangle&=\frac{v_0\sin{\theta_0}}{D_\theta}(1-e^{-D_\theta t})
\end{split}
\label{avgx}
\end{equation}
 We get the MSDs of the $x$ and the $y$ components separately: 
\begin{equation}
\begin{split}
\langle\Delta x^2(t)\rangle&=\langle x^2(t)\rangle-\langle x(t)\rangle^2\\
&=\Bigg[2\bar{D}t+\Delta D\cos{2\theta_0}\Big(\frac{1-e^{-4D_\theta t}}{4D_\theta}\Big)\Bigg]+\frac{v_0^2\cos{2\theta_0}}{12D_\theta^2}\big(3-4e^{-D_\theta t}+e^{-4D_\theta t}\big)+\frac{v_0^2}{D_\theta^2}\big(D_\theta t+e^{-D_\theta t}-1\big)\\
&-\frac{v_0^2\cos^2{\theta_0}}{D_\theta^2}(1-e^{-D_\theta t})^2
\end{split}
\label{varx}
\end{equation}
\begin{equation}
\begin{split}
\langle\Delta y^2(t)\rangle&=\langle y^2(t)\rangle-\langle y(t)\rangle^2\\
&=\Bigg[2\bar{D}t-\Delta D\cos{2\theta_0}\Big(\frac{1-e^{-4D_\theta t}}{4D_\theta}\Big)\Bigg]-\frac{v_0^2\cos{2\theta_0}}{12D_\theta^2}\big(3-4e^{-D_\theta t}+e^{-4D_\theta t}\big)+\frac{v_0^2}{D_\theta^2}\big(D_\theta t+e^{-D_\theta t}-1\big)\\
&-\frac{v_0^2\sin^2{\theta_0}}{D_\theta^2}(1-e^{-D_\theta t})^2
\end{split}
\label{vary}
\end{equation}
As our study is based on keeping initial orientational angle fixed at $\theta_0=0$, so $\langle y(t)\rangle=0$ and $\langle x(t)\rangle=\frac{v_0}{D_\theta}(1-e^{-D_\theta t})$. Therefore, $\langle\Delta x^2(t)\rangle_{\theta_0=0}=\langle x^2(t)\rangle_{\theta_0=0}-\frac{v_0^2}{D_\theta^2}(1-e^{-D_\theta t})^2$ and $\langle\Delta y^2(t)\rangle_{\theta_0=0}=\langle y^2(t)\rangle_{\theta_0=0}$.

\begin{figure}[h]
	    \centering
	     \includegraphics[width=1.05\textwidth]{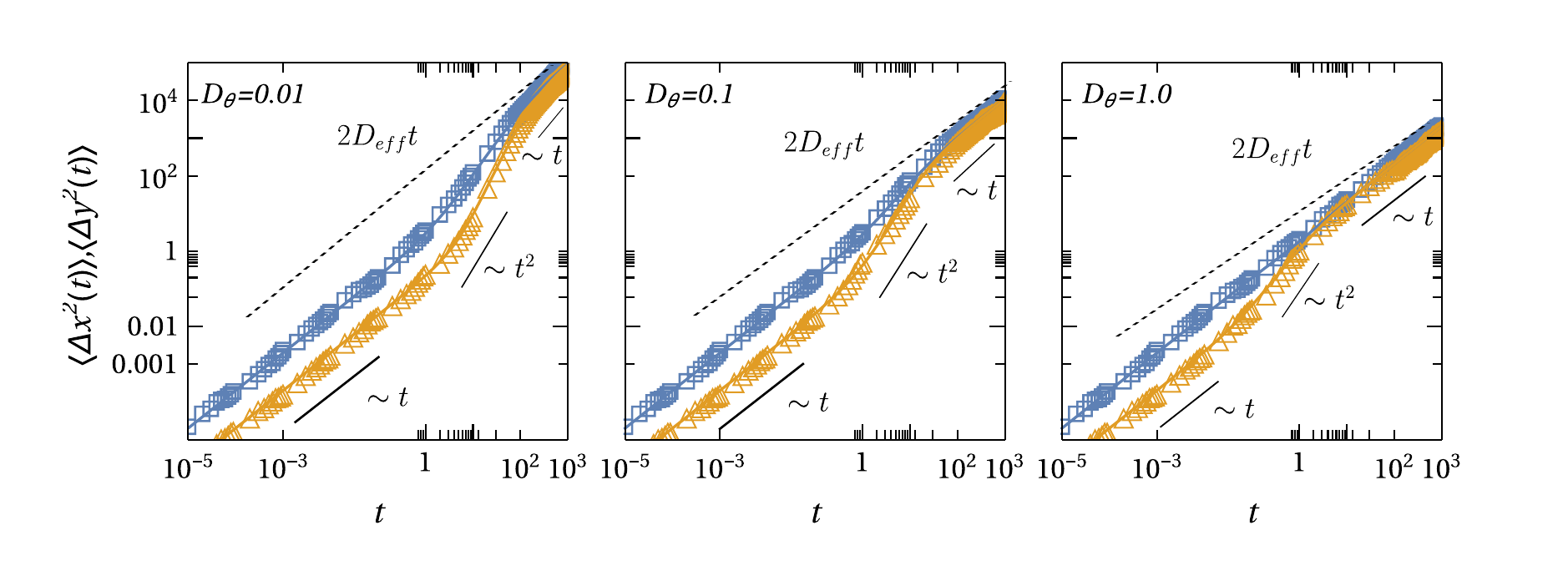}
	     \caption{Plot of mean-square displacement(MSD) along the $x$-direction (blue), $y$-direction(orange) with time for different rotational diffusivities $D_\theta$ as written on the figure. The translational diffusivities are fixed at $D_{\parallel}=1.0$, $D_{\perp}=0.1$. The propulsion velocity $v_0$ and the initial angle $\theta_0$ are fixed at $v_0=1$ and $\theta_0=0$, respectively. The solid lines are to express analytical results in Eq.(\ref{varx}), (\ref{vary}). The dashed lines are the representation of MSD at long time regime, which is purely diffusive.}
	     \label{msd_v1}
	    \end{figure}
     \begin{figure}[h]
	    \centering
	     \includegraphics[width=1.05\textwidth]{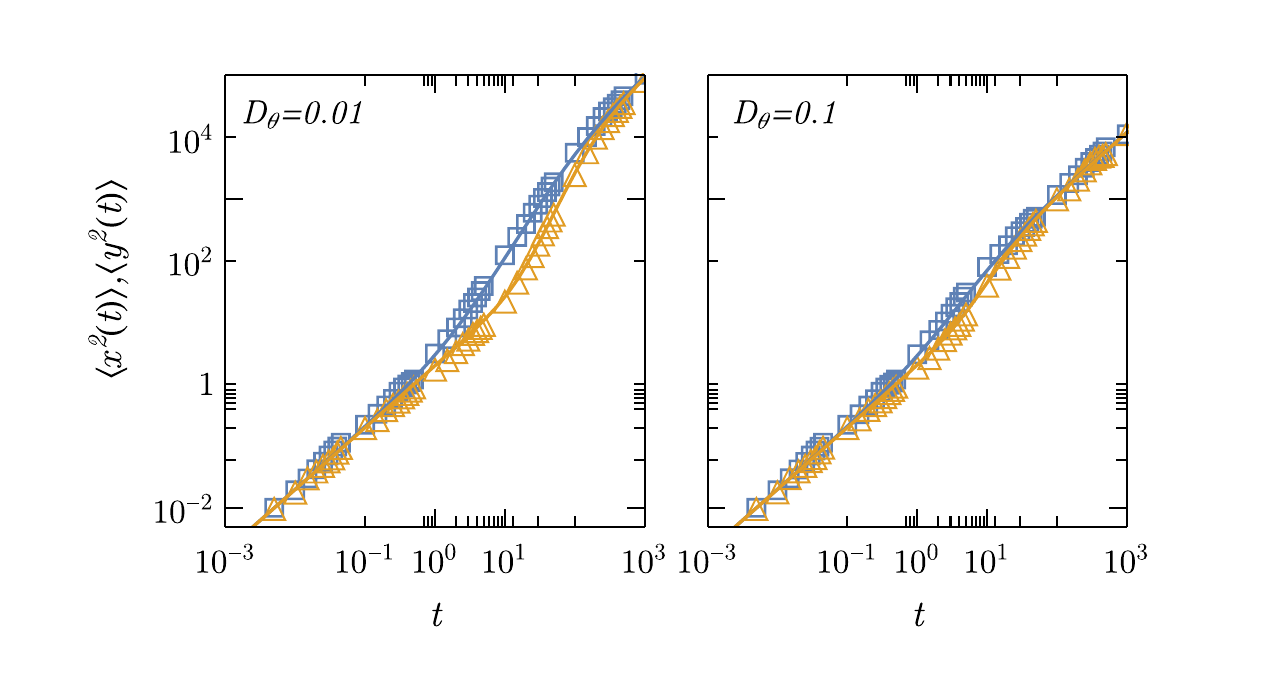}
	     \caption{Plot of mean-square displacement(MSD) for isotropic particle along the $x$-direction (blue), $y$-direction(orange) with time for different rotational diffusivities $D_\theta$ as written on the figure. The translational diffusivity is kept fixed at $\Bar{D}=1.0$. The propulsion velocity $v_0$ and the initial angle $\theta_0$ are fixed at $v_0=1$ and $\theta_0=0$, respectively. The solid lines are to express analytical results in Eq.(\ref{eqn:msd1}), (\ref{eqn:msd2}) considering $\Delta D=0$.}
	     \label{isotropic}
	    \end{figure}
 If we look at the results for isotropic particle, we can only take $\Delta D=0$ in Eqs.(\ref{eqn:msd1}) and (\ref{eqn:msd2}), which gives the results found in Hagen et al.\cite{ten2011brownian}. For an asymmetric particle the additional $\theta_0$ dependent term arises at very short time regime, which is not visible for the isotropic particle. The relative orientation of the linear channel's direction and the ellipsoidal particle's initial long axis direction is represented by this additional term.

\subsubsection{Short time regime $t<D_\theta^{-1}$}
 Let us consider the ellipsoidal ABP is starting from the origin $x=y=0$ initially with a given orientation, which
we choose to be along x-axis, so that $\theta_0=0$. For $t\ll D_\theta^{-1}$, if we expand the exponential terms of Eqs.(\ref{varx}) and (\ref{vary}) up to second order approximation, we find,

\begin{equation}
\begin{split}
&\langle\Delta x^2(t)\rangle=2D_{\parallel}t-2D_\theta\Delta Dt^2\\
&\langle\Delta y^2(t)\rangle=2D_{\perp}t+2D_\theta\Delta D t^2
\end{split}
\label{small_D}
\end{equation}
From Eq.(\ref{small_D}), it is evident that in the transient time regime, the $t^2$ term can be neglected, and the term with $t$ will contribute. Consequently, the initial mean square displacements (MSDs) along the $x$ and $y$ axes are $2D_{\parallel} t$ and $2D_{\perp}t$, respectively. For a symmetric particle in the $t\ll D_\theta^{-1}$ regime, we can perform an expansion of the exponential terms in Eqs.(\ref{eqn:msd1}) and (\ref{eqn:msd2}), setting $\Delta D=0$ up to the first order. This yields $\langle x^2(t)\rangle_{t\ll D_\theta^{-1}}=\langle y^2(t)\rangle_{t\ll D_\theta^{-1}}=2\bar{D}t$, where $\bar{D}$ is the translational diffusion coefficient of an isotropic particle. Consequently, it can be inferred that for an asymmetric particle in the $t\ll D_\theta^{-1}$ regime, the initial mean square displacements (MSDs) along the $x$ and $y$ axes differ, whereas for an isotropic particle, both are the same, the simulation results are shown in the fig.(\ref{isotropic}). In this regime, we observe simple diffusive behavior along both axes related to their translational diffusion coefficients, respectively. As time evolves, Eq.(\ref{small_D}) is predominantly influenced by the term involving $t^2$, consequently leading to the emergence of the super-diffusive regime. Our analytical and simulation results are showcased in fig.(\ref{msd_v1}), portraying a transient diffusive regime succeeded by a super-diffusive regime until $t\approx D_\theta^{-1}$ for various values of $D_\theta$.

\begin{figure}[h]
	    \centering
	     \includegraphics[width=0.9\textwidth]{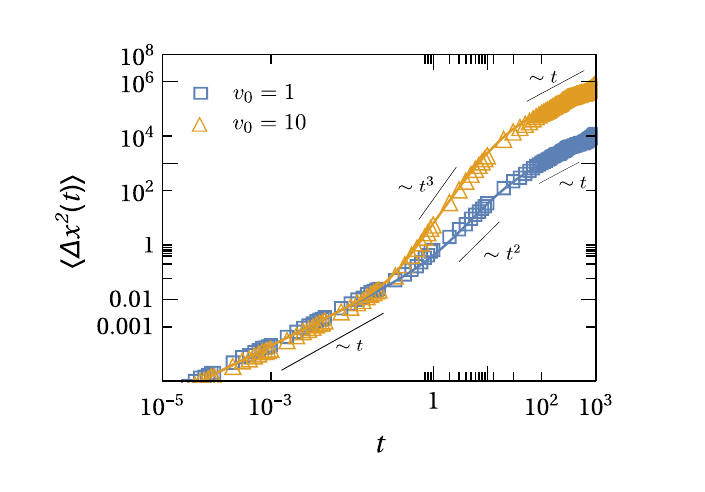}
	     \caption{Plot of mean-square displacement (MSD) along the $x$-direction with time for different propulsion velocities $v_0$ as written on the figure. The translational diffusivities are fixed at $D_{\parallel}=1.0$, $D_{\perp}=0.1$. The rotational diffusivity and the initial angle $\theta_0$ are fixed at $D_\theta=0.1$ and $\theta_0=\pi/2$ respectively. The solid lines are to express analytical results in Eq.(\ref{eqn:msd1}).}
	     \label{fig:msd_v1}
	    \end{figure}

 If we change the initial orientational angle to $\theta_0=\pi/2$, which implies that the MSD along $x$ and $y$-axes are interchanged and observe the effect of changing propulsion velocity in fig.(\ref{fig:msd_v1}). We find that the mean-square displacement shows super- ballistic power-law behavior for large propulsion velocity. If we expand the Eq.(\ref{eqn:msd1}) putting $\theta_0=\pi/2$ (for $\theta_0=\pi/2$, $\langle x(t)\rangle$ of Eq.(\ref{avgx}) becomes zero, so $\langle\Delta x^2(t)\rangle_{\theta_0=\pi/2}=\langle x^2(t)\rangle_{\theta_0=\pi/2}$) upto third order then we get 
 \begin{equation}
 \langle\Delta x^2(t)\rangle_{\theta_0=\pi/2}=2D_\parallel t+2D_\theta\Delta Dt^2+\frac{2}{3}(v_0^2D_\theta-4\Delta D D_\theta^2)t^3
 \label{t3}
 \end{equation}
 In the given expression, it is notably observed that as the value of $v_0$ increases, the coefficient of $t^3$ containing $v_0^2$ becomes increasingly dominant. This observation indicates a super- ballistic power-law behavior associated with the system in fig.(\ref{fig:msd_v1}) in the intermediate regime. Hence, the nature of the crossover to an intermediate super-diffusive regime is contingent on both the initial orientational angle $\theta_0$ and the effective propulsion velocity $v_0$. When the initial orientation possesses a significantly large component of propulsion velocity parallel to the $x$ axis, the mean-square displacement undergoes a transition to a ballistic regime, dictated by a scaling relation proportional to $t^2$. Conversely, when the initial orientation of the particle is substantial, aligning with $\pi/2$, meaning that the major axis of the particle points along the $y$ axis initially, the particle experiences angular displacement through rotational diffusion. This results in an increased projected propulsion velocity towards the $x$ axis. Only under these conditions does the velocity induce a lateral displacement that matches or exceeds the displacements caused by the original translational motion. Because of the multiplicative coupling between diffusive and ballistic behaviors in angular and translational displacements, respectively, the mean-square displacement exhibits a super-ballistic power-law behavior. Despite the absence of apparent acceleration in the system, the velocity along the $x$ direction changes proportionally to the projection of the velocity onto the $x$ axis induced by rotational Brownian motion. Transient hyperdiffusion has also been detected in generalized Brownian motion within a tilted washboard potential, where it was attributed to the transient heating of particles by the thermal bath\cite{siegle2010origin}. In the underdamped active fractional Langevin equation, the interaction between active noise and long-time viscoelastic memory effects results in unique and intricate nonequilibrium hyperdiffusion dynamics\cite{joo2023viscoelastic}.

If we calculate the two dimensional translational motion we can find from the expression $\langle r^2(t)\rangle=\langle x^2(t)\rangle+\langle y^2(t)\rangle$. The exact expression can be found from Eq.(\ref{eqn:msd1}) and Eq.(\ref{eqn:msd2}) as,
\begin{equation}
    \langle r^2(t)\rangle=4\Bar{D}t+\frac{2v_0^2}{D_\theta^2}(e^{-D_\theta t}+D_\theta t-1)
\end{equation}
In this expression, the dependence on $\theta_0$ vanishes, similar to isotropic particles. The representation of diffusive motion is succinctly captured by the term $4\Bar{D}t$. 

In the regime where time extends significantly, that is, when $t\gg D_\theta^{-1}$, the exponential components of the Eqs.(\ref{varx}) and (\ref{vary}) become negligible. Consequently, the mean-square displacements along both the $x$ and $y$ axes exhibit diffusive behavior similar to an isotropic particle expressed as,
\begin{equation}
\begin{split}
    &\langle\Delta x^2(t)\rangle=2(\Bar{D}+\frac{v_0^2}{2D_\theta})t+\frac{\Delta D}{4D_\theta}-\frac{7v_0^2}{4D_\theta^2}\\
    &\langle\Delta y^2(t)\rangle=2(\Bar{D}+\frac{v_0^2}{2D_\theta})t-\frac{\Delta D}{4D_\theta}-\frac{5v_0^2}{4D_\theta^2}
\end{split}
\label{msdlong}
\end{equation}
The effective diffusion coefficient, denoted as $D_{eff}=\Bar{D}+\frac{v_0^2}{2D_\theta}$. For an isotropic particle also Eq.(\ref{msdlong}) is valid with $\Delta D=0$, indicating a constant effective diffusivity $D_{eff}=\Bar{D}+\frac{v_0^2}{2D_\theta}$, where $\Bar{D}$ stands for the translational diffusion coefficient of an isotropic particle. This relationship is visually represented by the dashed line in fig.(\ref{msd_v1}). In the context of anisotropic ABP, the evolving dynamics can be delineated into three distinct regions. The initial region exhibits purely diffusive behavior, characterized by its transient nature. The second region displays super-diffusive characteristics, contingent upon the propulsion velocity. Lastly, the third region reverts to diffusive dynamics.

\subsubsection{Diffusion coefficients}
An anisotropic particle, initially oriented at a specific angle, exhibits faster diffusion along its long axis compared to its short axis. However, as time advances, memory of initial directional fades away and leads to a crossover from anisotropic diffusion at short times to isotropic diffusion at times much longer than the time scale $D_\theta^{-1}$. We will compute three translational diffusion coefficients: $D_{11}$ for motion in the $x$ direction, $D_{22}$ for motion in the $y$ direction, and the cross-diffusion coefficient $(D_{12})$, defined as

\begin{equation}
\begin{split}
D_{ij}(t)=\frac{\langle \Delta x_i(t) \Delta x_j(t)\rangle_{\theta_{0}}^{\eta_1,\eta_2,\eta_3}-\langle \Delta x_i(t)\rangle_{\theta_0}^{\eta_1,\eta_2,\eta_3} \cdot \langle \Delta x_j(t) \rangle_{\theta_0}^{\eta_1,\eta_2,\eta_3}}{2t}
\end{split}
\label{eqn:diff_xy}
\end{equation}
where $(x_1, x_2)=(x, y)$. Throughout the remainder of the paper, we will omit superscripts and subscripts, unless the average is performed over a particular noise source rather than over all three.



Using, Eqs.({\ref{eqn:msd1}), (\ref{eqn:msd2}), and (\ref{eqn:diff_xy}) we get,

\begin{equation}
\begin{split}
D_{11}(t)=\bar{D}+\frac{\Delta D}{2t}\cos 2\theta_0\Big(\frac{1-e^{-4D_\theta t}}{4D_\theta}\Big)+\frac{v_0^2\cos 2\theta_0}{24D_\theta^2t}\big(3-4e^{-D_\theta t}+e^{-4D_\theta t}\big)+\frac{v_0^2}{2D_\theta^2t}\big(D_\theta t+e^{-D_\theta t}-1\big)
\end{split}
\label{eqn:diff_x}
\end{equation}

\begin{equation}
\begin{split}
D_{22}(t)=\bar{D}-\frac{\Delta D}{2t}\cos 2\theta_0\Big(\frac{1-e^{-4D_\theta t}}{4D_\theta}\Big)-\frac{v_0^2\cos 2\theta_0}{24D_\theta^2t}\Big(3-4e^{-D_\theta t}+e^{-4D_\theta t}\Big)+\frac{v_0^2}{2D_\theta^2t}\Big(D_\theta t+e^{-D_\theta t}-1\Big)
\end{split}
\label{eqn:diff_y}
\end{equation}	 

To calculate the cross diffusion terms we proceed as,
\begin{equation}
\begin{split}
D_{12}(t)&=\frac{1}{2t}\Big\langle\Big[\int_{0}^{t}v_0\cos{\theta(t^\prime)}dt^\prime+\int_{0}^{t}\eta_1(t^\prime)dt^\prime\Big]\Big[\int_{0}^{t}v_0\sin{\theta(t^{\prime\prime})}dt^{\prime\prime}+\int_{0}^{t}\eta_2(t^{\prime\prime})dt^{\prime\prime}\Big]\Big\rangle\\
&=\frac{1}{2t}\Big[v_0^2\int_{0}^{t}dt^\prime\int_{0}^{t}dt^{\prime\prime}\langle\cos{\theta(t^\prime)}\sin{\theta(t^{\prime\prime})}\rangle+\int_{0}^{t}dt^\prime\int_{0}^{t}dt^{\prime\prime}\langle\eta_1(t^\prime)\eta_2(t^{\prime\prime})\rangle\Big]\\
&=\frac{1}{2t}\Bigg[\frac{v_0^2}{2}\int_{0}^{t}dt^\prime\int_{0}^{t}dt^{\prime\prime}\Big(\sin{2\theta_0}e^{-D_\theta(t^\prime+t^{\prime\prime}+2\min(t^\prime,t^{\prime\prime}))}-e^{-D_\theta(t^\prime+t^{\prime\prime}-2\min(t^\prime,t^{\prime\prime}))}\Big)+k_BT\Delta\Gamma\sin{2\theta_0}\int_{0}^{t}dt^\prime e^{-4D_\theta t^\prime}\Bigg]\\
&=\frac{v_0^2\sin{2\theta_0}}{24D_\theta^2t}(3+e^{-4D_\theta t}-4e^{-D_\theta t})-\frac{v_0^2}{2D_\theta^2t}[e^{-D_\theta t}+D_\theta t-1]+\frac{\Delta D\sin{2\theta_0}}{8D_\theta t}(1-e^{-4D_\theta t})
\end{split}
\label{cross}
\end{equation}
The detailed calculations of the above integrals have been shown in the Eq.(\ref{I34}).

For a passive Brownian particle propulsion velocity $v_0=0$, so the above Eqs.(\ref{eqn:diff_x}), (\ref{eqn:diff_y}) and (\ref{cross}) can be written as,
\begin{equation}
\begin{split}
&D_{11}(t)=\bar{D}+\frac{\Delta D}{2t}\cos 2\theta_0\Big(\frac{1-e^{-4D_\theta t}}{4D_\theta}\Big)\\
\end{split}
\label{eqn:diff_f1}
\end{equation}

\begin{equation}
\begin{split}
&D_{22}(t)=\bar{D}-\frac{\Delta D}{2t}\cos 2\theta_0\Big(\frac{1-e^{-4D_\theta t}}{4D_\theta}\Big)\\
\end{split}
\label{eqn:diff_f2}
\end{equation}
\begin{equation}
\begin{split}
D_{12}(t)=\frac{\Delta D\sin{2\theta_0}}{8D_\theta t}(1-e^{-4D_\theta t})
\end{split}
\label{crossd}
\end{equation}
Here it can be seen that the long time diffusion coefficients $D_{11}$, $D_{22}$ and $D_{12}$ match the results found analytically\cite{grima2007brownian} and experimentally\cite{han2006brownian}. Based on experimental results, it has been confirmed that a freely moving ellipsoidal micro-sized particle with an aspect ratio of $1:8$ exhibits a transient decay time scale in the range of seconds\cite{han2006brownian}, but particles having a propulsion velocity show different behavior.

The temporal variation of the diffusion coefficients in the early time regime is attributed to the anomalous diffusion behavior induced by both the propulsion velocity and the particle's anisotropy. As depicted in fig.(\ref{fig56}) during the transient time regime, the diffusion coefficient initially maintains a constant value for a very brief period. However, as time progresses, a time-dependent behavior emerges. This is because, in the transient time regime, the particle's dynamics exhibit diffusive behavior for an extremely short duration. Subsequently, as time evolves, the particle's dynamics shift towards super-diffusion until $t\approx D_\theta^{-1}$, leading to a time dependency of diffusion coefficients. Furthermore, in the long time regime, the dynamics revert to purely diffusive behavior, characterized by an effective diffusion coefficient $D_{eff}$.

\begin{figure}
  \centering
  \subfigure[Variation of diffusion coefficients along the $x$ ($D_{11}$ represented by blue square) and $y$-directions ($D_{22}$ represented by orange triangle) with time for different $D_\theta$ values indicated in the figure and propulsion velocity is kept fixed at $v_0=1$.]{\label{fig:Diff_D_time_v2}\includegraphics[width=0.495\textwidth]{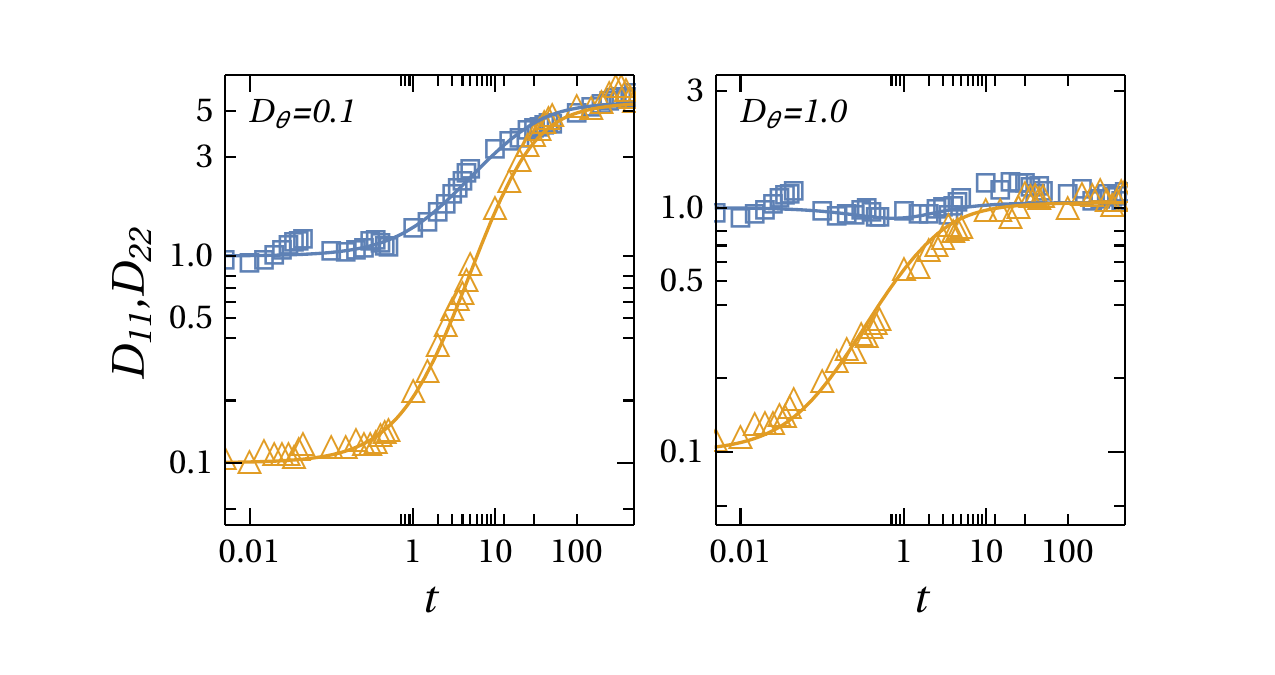}}
  \hfill
  \subfigure[Variation of diffusion coefficients along the $x$ ($D_{11}$ represented by open blue square) and $y$-directions ($D_{22}$ represented by open orange triangle) with time for different $v_0$ values indicated in the figure and the rotational diffusion coefficient is kept fixed at $D_\theta=1$.]{\label{fig:Diff_D_time_v}\includegraphics[width=0.495\textwidth]{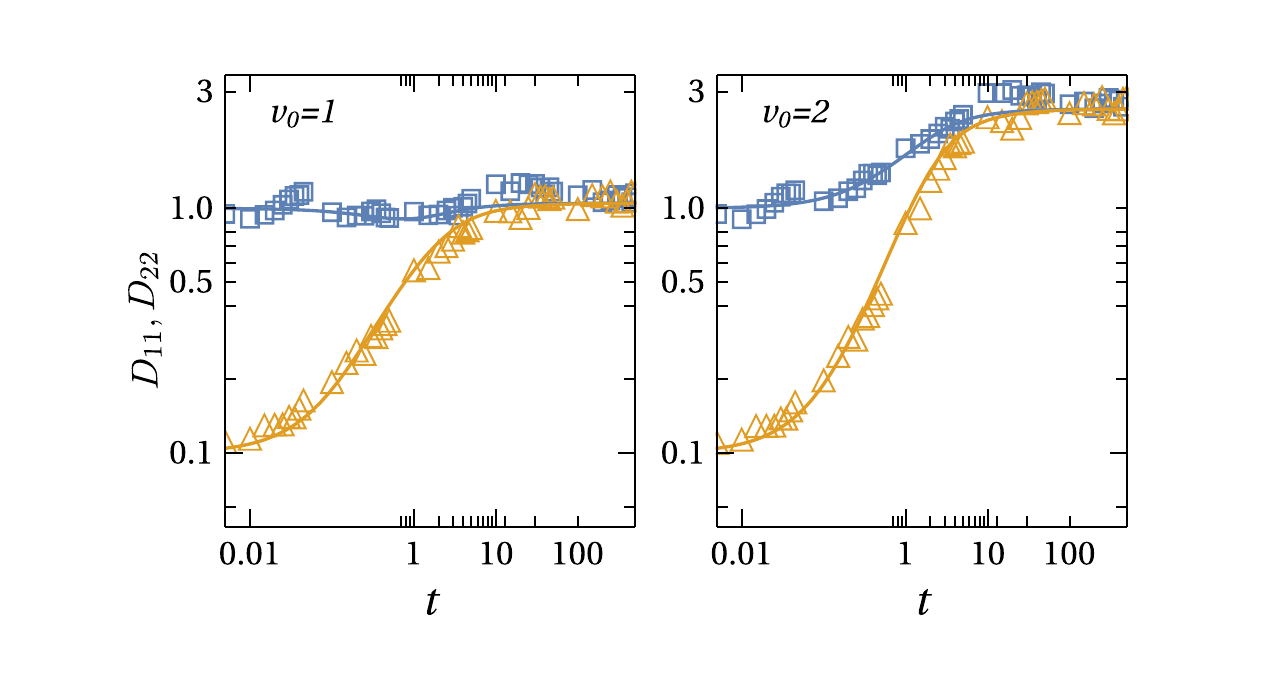}}
  \caption{Diffusion coefficients along $x$ and $y$-directions are respectively $D_{11}$(open blue square) and $D_{22}$(open orange triangle). The translational diffusivities are fixed at $D_{\parallel}=1.0$, $D_{\perp}=0.1$ and the initial orientational angle $\theta_0=0$. The solid lines are to express analytical results in Eq.(\ref{eqn:diff_x}), (\ref{eqn:diff_y}).}
  \label{fig56}
\end{figure}


For ABP propulsion velocity $v_0$ has a finite value. Therefore, we can see from Eq.(\ref{eqn:diff_x}) and (\ref{eqn:diff_y}) the third and fourth terms consist $v_0$, so these terms will also additionally contribute. At $t \ll D_\theta^{-1}$, Eq.(\ref{eqn:diff_x}) and (\ref{eqn:diff_y}) become $D_{11}=D_\parallel$ and $D_{22}=D_\perp$ as we expand the exponential terms up to first order for the initial orientational angle $\theta_0=0$. In fig.(\ref{fig:Diff_D_time_v2}) we have shown the time variation of diffusion coefficients for an active ellipsoidal particle. These results match with our analytical results. In fig.(\ref{fig:Diff_D_time_v}) we can see how diffusion coefficients vary with time for different values of propulsion velocity $v_0$. We can see from the fig.(\ref{fig:Diff_D_time_v}) that with $v_0=2$, the variation of the diffusion coefficient is much greater than the lower propulsion velocities.

In longer time limit we can see from fig.(\ref{fig:Diff_D_time_v2}) and fig.(\ref{fig:Diff_D_time_v}) that, $D_{11}$ and $D_{22}$ reach a time-independent constant value. This can be shown from Eq.(\ref{eqn:diff_x}, \ref{eqn:diff_y}) putting the condition that $t$ is very large. The second and the third terms in Eqs.(\ref{eqn:diff_x}) and (\ref{eqn:diff_y}) consisting $t^{-1}$ disappear for large $t$. The fourth term with the coefficient $\frac{v_0^2}{D_\theta^2}$ gives $t$ independent value only for the term clubbing with $D_\theta t$, as in that case $t$ cancels out. But the other two terms disappear. Thus we get the value of $D_{11}$ and $D_{22}$ as time independent constants for a set of different mobility parameters. This time-independent value can be defined as the effective diffusion coefficient $D_{eff}$ which is also found from Eq.(\ref{msdlong}),
\begin{equation}
    D_{eff}=\Bar{D}+\frac{v_0^2}{2D_\theta}
    \label{eqn:diff_free1}
\end{equation}

From the numerical results in fig.(\ref{fig:Diff_D_time_v2}), we get the value of $D_{11}=D_{22}\approx 5.55$ for $D_\theta=0.1$ at longer time. From Eq.(\ref{eqn:diff_free1}), we get the analytical value $D_{eff}=5.55$ and for $D_\theta=1.0$, $D_{eff}=1.05$, which also matches with the numerical results of \cref{fig:Diff_D_time_v2}. From these results we can see that for the smaller $D_\theta$ values, the directional transport($D_{11}$, $D_{22}$) is higher, and as gradually $D_\theta$ value increases, directional transport decreases.

From fig.(\ref{fig:Diff_D_time_v}) it can be seen that, analytical values of $D_{eff}$ from Eq.(\ref{eqn:diff_free1}) $D_{eff}=1.05$ for $v_0=1.0$ and $D_{eff}=2.55$ for $v_0=2.0$ exactly match with the numerical values of $D_{11}$, $D_{22}$ at longer time limit.


\subsection{Harmonically trapped system}
Since laser traps are typically used in experiments to monitor colloidal particles, it is important to talk about the case in which a harmonic trap captures an active ellipsoidal particle. Following, we assume that there is no favored alignment direction and that the harmonic trap is isotropic. Furthermore, if we assume strong confinement, the particle's deviation from its mean position is essentially zero in late times. The angular displacements so obey Gaussian statistics as the particle rotates freely. The potential confinement has the form $U(x, y) = \kappa(x^2+y^2)/2$, where $\kappa$ is the stiffness coefficient of the trap. The corresponding Langevin equation from Eq.(\ref{lab_frame}) takes the form

\begin{equation}
\begin{split}
&\frac{\partial x}{\partial t}=v_0\cos\theta-\kappa x(\bar{\Gamma}+\frac{1}{2}\Delta\Gamma\cos\theta(t))-\frac{1}{2}\kappa y\Delta\Gamma\sin\theta(t)+\eta_1(t)\\
&\frac{\partial y}{\partial t}=v_0\sin\theta-\frac{1}{2}\kappa x\Delta\Gamma\sin\theta(t)-\kappa y(\bar{\Gamma}-\frac{1}{2}\Delta\Gamma\cos\theta(t))+\eta_2(t)\\
&\frac{\partial \theta}{\partial t}=\Gamma_{\theta}\tau(t)+\eta_3(t)
\end{split}
\label{Langevin_har}
\end{equation}
 According to the framework described in Eq.(\ref{torque}), the magnitude of the torque is represented by $\tau=xF_y-yF_x$. In the scenario where the particle is constrained within a symmetric harmonic trap, as considered in our model ($F_x=-\kappa x$ and $F_y=-\kappa y$), it is observed that the magnitude of the torque becomes zero. Consequently, the rotational dynamics of the particle within the confinement of a symmetric harmonic potential evolves solely based on fluctuations, thereby demonstrating a purely diffusive rotational behavior. 

By looking at Eq.(\ref{Langevin_har}), we find that in absence of any asymmetry the equations reduce to that of an isotropic particle and the correction due to the shape asymmetry comes in the combination of $\kappa\Delta\Gamma/2$. Furthermore, the equations of motion in Eq.(\ref{Langevin_har}) are coupled and therefore these are non-Markovian in behavior. The coupling vanishes in the limit of weak anisotropy of $\Delta\Gamma\rightarrow 0$ since it is proportional to the difference in the mobilities $\Delta\Gamma$. In this problem, we will assume weak asymmetry. Let us define Eq.(\ref{Langevin_har}) in general form as
\begin{equation}
    \Dot{\bold{R}}=-\kappa[\Bar{\Gamma}\mathbf{1}+\frac{\Delta\Gamma}{2}\Bar{\Bar{\mathcal{R}}}(t)]\bold{R}(t)+v_0\bold{\hat{n}}+\eta(t)
    \label{21123}
\end{equation}
In this problem, we will assume weak asymmetry. To solve this equation we take the perturbative expansion, 
\begin{equation}
    \bold{R}(t)=\bold{R}_0(t)-(\frac{\kappa\Delta\Gamma}{2})\bold{R}_1(t)+(\frac{\kappa\Delta\Gamma}{2})^2\bold{R}_2(t)+\mathcal{O}(\frac{\kappa\Delta\Gamma}{2})^3
    \label{22123}
\end{equation}
Substituting Eq.(\ref{22123}) in Eq.(\ref{21123}) and equalizing both sides we get the equations for $\bold{R}_0(t)$ and $\bold{R}_1(t)$ as
\begin{equation}
    \begin{split}
        &\Dot{\bold{R}}_0(t)=-\kappa\Bar{\Gamma}\bold{R}_0(t)+v_0\bold{\hat{n}}(t)+\eta(t)\\
        &\Dot{\bold{R}}_1(t)=-\kappa\Bar{\Gamma}\bold{R}_1(t)+\Bar{\Bar{\mathcal{R}}}(t)\bold{R}_0(t)\\
        &\Dot{\bold{R}}_2(t)=-\kappa\Bar{\Gamma}\bold{R}_2(t)+\Bar{\Bar{\mathcal{R}}}(t)\bold{R}_1(t)
    \end{split}
    \label{23123}
\end{equation}
The solutions for Eq.(\ref{23123}) defining the initial condition $\bold{R}(0)=0$, becomes
\begin{equation}
    \begin{split}
        &\bold{R}_0(t)=\int_{0}^{t}dt^\prime e^{-\kappa\Bar{\Gamma}(t-t^\prime)}\big[\eta(t^\prime)+v_0\bold{\hat{n}}(t^\prime)\big]\\
        &\bold{R}_1(t)=\int_{0}^{t}dt^\prime e^{-\kappa\Bar{\Gamma}(t-t^\prime)}\Bar{\Bar{\mathcal{R}}}(t^\prime)\bold{R}_0(t^\prime)\\
        &\bold{R}_2(t)=\int_{0}^{t}dt^\prime e^{-\kappa\Bar{\Gamma}(t-t^\prime)}\Bar{\Bar{\mathcal{R}}}(t^\prime)\bold{R}_1(t^\prime)
    \end{split}
    \label{24}
\end{equation}
The explicit form of the correlation matrix $\langle R_i(t)R_j(t)\rangle$ in the equal time, is given by
\begin{equation}
    \begin{split}
    \langle R_i(t)R_j(t)\rangle_{\eta,\theta}&=\langle R_{0,i}(t)R_{o,j}(t)\rangle_{\eta,\theta}-(\frac{\kappa\Delta\Gamma}{2})\langle R_{0,i}(t)R_{1,j}\rangle_{\eta,\theta}+(\frac{\kappa\Delta\Gamma}{2})^2\Big[\langle R_{1,i}(t)R_{1,j}(t)\rangle_{\eta,\theta}\\
    &+2\langle R_{0,i}(t)R_{2,j}(t)\rangle_{\eta,\theta}\Big]+\mathcal{O}(\frac{\kappa\Delta\Gamma}{2})^3
    \end{split}
    \label{25}
\end{equation}
Here we have considered the fact that $\langle R_{0,i}R_{1,j}\rangle=\langle R_{0,j}R_{1,i}\rangle$. 
We now start to calculate the different terms of the correlation matrix. The correlation matrix for $\bold{R}_0(t)$ is given as averaging over the translational and the rotational noise. Now considering $i=j=x$, we can calculate the values of $\langle x_0^2(t)\rangle$ and $\langle x_0(t)x_1(t)\rangle$ of Eq.(\ref{25}) shown by Ghosh et al.\cite{ghosh2022persistence}.
\begin{equation}
    \begin{split}
        \langle x_0^2(t)\rangle&=\frac{k_BT}{\kappa}(1-e^{-2\kappa\Bar{\Gamma}t})+k_BT\Delta\Gamma\cos{2\theta_0}\Bigg(\frac{e^{-4D_\theta t}-e^{-2\kappa\Bar{\Gamma}t}}{2\kappa\Bar{\Gamma}-4D_\theta}\Bigg)+\frac{v_0^2\cos{2\theta_0}}{2}\Bigg[\frac{D_\theta(e^{-4D_\theta t}-e^{-2\kappa\Bar{\Gamma}t})}{(\kappa\Bar{\Gamma}-D_\theta)(\kappa\Bar{\Gamma}-2D_\theta)(\kappa\Bar{\Gamma}-3D_\theta)}\\
        &+\frac{e^{-4D_\theta t}-2e^{-(\kappa\Bar{\Gamma}+D_\theta)t}+e^{-2\kappa\Bar{\Gamma}t}}{(\kappa\Bar{\Gamma}-D_\theta)(\kappa\Bar{\Gamma}-3D_\theta)}\Bigg]+\frac{v_0^2}{2}\Bigg[\frac{1-2e^{-(\kappa\Bar{\Gamma}+D_\theta)t}+e^{-2\kappa\Bar{\Gamma}t}}{(\kappa\Bar{\Gamma}-D_\theta)(\kappa\Bar{\Gamma}+D_\theta)}-\frac{D_\theta(1-e^{-2\kappa\Bar{\Gamma}t})}{\kappa\Bar{\Gamma}(\kappa\Bar{\Gamma}-D_\theta)(\kappa\Bar{\Gamma}+D_\theta)}\Bigg]
    \end{split}
    \label{2645}
\end{equation}
\begin{equation}
    \begin{split}
        \langle x_0(t)x_1(t)\rangle&=\frac{k_BT}{\kappa}\cos{2\theta_0}\Bigg[\frac{e^{-4D_\theta t}-e^{-2\kappa\Bar{\Gamma}t}}{2\kappa\Bar{\Gamma}-4D_\theta}-\frac{e^{-2\kappa\Bar{\Gamma}t}-e^{-(2\kappa\Bar{\Gamma}+4D_\theta)t}}{4D_\theta}\Bigg]+\frac{k_BT}{\kappa}\frac{\Delta\Gamma}{2\Bar{\Gamma}}\Bigg[\frac{1-e^{-2\kappa\Bar{\Gamma}t}}{2\kappa\Bar{\Gamma}+4D_\theta}\\
        &-\frac{\kappa\Bar{\Gamma}}{2D_\theta}\frac{e^{-2\kappa\Bar{\Gamma}t}-e^{-(2\kappa\Bar{\Gamma}+4D_\theta)t}}{\kappa\Bar{\Gamma}+4D_\theta}\Bigg]-\frac{3v_0^2D_\theta e^{-\kappa\Bar{\Gamma}t}\sinh{\kappa\Bar{\Gamma}t}}{2\kappa\Bar{\Gamma}(\kappa\Bar{\Gamma}+D_\theta)(\kappa\Bar{\Gamma}-D_\theta)(\kappa\Bar{\Gamma}+2D_\theta)}
    \end{split}
    \label{2646}
\end{equation}
Now we have considered up to the first order correction for Eq.(\ref{25}) and thus by using above Eqs.(\ref{2645}) and (\ref{2646})  the expression for mean-square displacement along the $x$ axis is given by

\onecolumngrid

\begin{widetext}
\begin{equation}
    \begin{split}
        \langle x^2(t)\rangle&=\frac{k_BT}{\kappa}(1-e^{-2\kappa\bar{\Gamma}t})+\frac{k_BT\Delta\Gamma}{4D_\theta}\Bigg[e^{-2\kappa\bar{\Gamma}t}-e^{-(2\kappa\Bar{\Gamma}+4D_\theta)t}+\kappa\Delta\Gamma\frac{e^{-2\kappa\bar{\Gamma}t}-e^{-(2\kappa\Bar{\Gamma}+4D_\theta)t}}{\kappa\Bar{\Gamma}+4D_\theta}\Bigg]\\
        &+\frac{v_0^2\cos{2\theta_0}}{2}\Bigg[\frac{D_\theta(e^{-4D_\theta t}-e^{-2\kappa\Bar{\Gamma}t})}{(\kappa\Bar{\Gamma}-D_\theta)(\kappa\Bar{\Gamma}-2D_\theta)(\kappa\Bar{\Gamma}-3D_\theta)}+\frac{e^{-4D_\theta t}-2e^{-(\kappa\Bar{\Gamma}+D_\theta)t}+e^{-2\kappa\Bar{\Gamma}t}}{(\kappa\Bar{\Gamma}-D_\theta)(\kappa\Bar{\Gamma}-3D_\theta)}\Bigg]\\
        &+\frac{v_0^2}{2}\Bigg[\frac{1-2e^{-(\kappa\Bar{\Gamma}+D_\theta)t}+e^{-2\kappa\Bar{\Gamma}t}}{(\kappa\Bar{\Gamma}-D_\theta)(\kappa\Bar{\Gamma}+D_\theta)}-\frac{D_\theta(1-e^{-2\kappa\Bar{\Gamma}t})}{\kappa\Bar{\Gamma}(\kappa\Bar{\Gamma}-D_\theta)(\kappa\Bar{\Gamma}+D_\theta)}\Bigg]\\
        &+(\kappa\Delta\Gamma)\sinh{\kappa\Bar{\Gamma} t}e^{-\kappa\Bar{\Gamma}t}\Bigg[\frac{3v_0^2D_\theta}{2\kappa\Bar{\Gamma}(\kappa\Bar{\Gamma}+D_\theta)(\kappa\Bar{\Gamma}-D_\theta)(\kappa\Bar{\Gamma}+2D_\theta)}-\frac{\Delta D}{2\kappa\Bar{\Gamma}(\kappa\Bar{\Gamma}+2D_\theta)}\Bigg]
    \end{split}
    \label{eqn:msdx_kappa}
\end{equation}
\end{widetext}


The calculation of MSD along $y$, $\langle y^2(t)\rangle$ has been shown in the Appendix(\ref{App}) and the final expression is shown in the Eq.(\ref{y2t}).
\begin{figure}
  \centering
  \subfigure[MSD along the $x$ and $y$ -directions in harmonic potential for different $\kappa$ values are indicated in the legend($\kappa=0.1$(open square), $\kappa=0.5$(open triangle), and $\kappa=1$(open circle)), here $D_\theta=1$ and $v_0=1$ are taken as fixed parameters. The solid lines for $\langle x^2(t)\rangle$ are the representations of the analytical expression of Eq.(\ref{eqn:msdx_kappa}), and the solid lines for $\langle y^2(t)\rangle$ are the representations of the analytical expression of Eq.(\ref{y2t}).]{\label{fig:msd_kappa1}\includegraphics[width=0.49\textwidth]{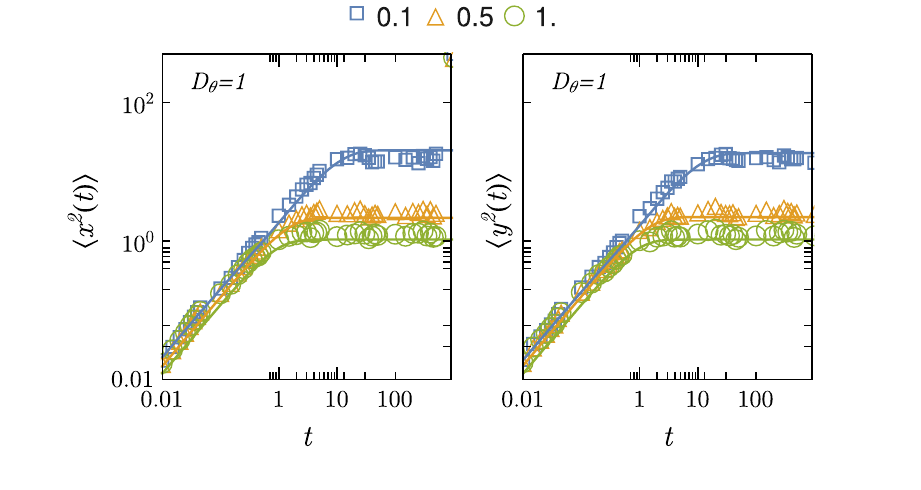}}
  \hfill
  \subfigure[MSD along the $x$-direction in harmonic potential for different choices of propulsion velocities of the anisotropic particle: $v_0=1$ (blue squares), $v_0=2$ (orange triangles), $v_0=3$ (green circles). Rotational diffusion coefficient and stiffness coefficient are kept fixed at $D_\theta=0.5$, and $\kappa=0.1$. The solid lines are to express analytical results in Eq.(\ref{eqn:msdx_kappa}).]{\label{fig:msd_kappa2}\includegraphics[width=0.49\textwidth]{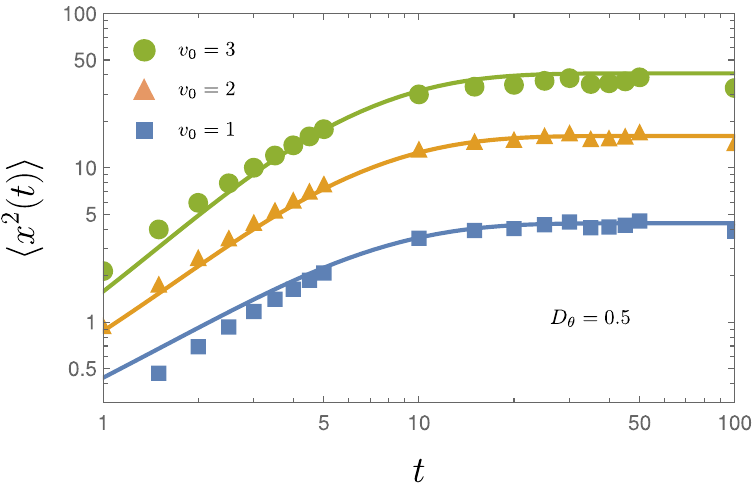}}
  \caption{The translational diffusivities are fixed at $D_{\parallel}=1.0$, $D_{\perp}=0.9$ and the initial angle $\theta_0$ is kept fixed at $\theta_0=0$.}
\end{figure}

At $\kappa \rightarrow 0$ (free particle) we get the Eq.(\ref{eqn:msd1}) from Eq.(\ref{eqn:msdx_kappa}).	
Fig.(\ref{fig:msd_kappa1}) displays the time variation of mean-square displacement along $x$ and $y$ axes in a harmonic trap for different $\kappa$ values as it represents the strength in the harmonic trap. As $\kappa$ increases, particle movement becomes more restricted, leading to a constrained rate of increment in the mean-square displacement value. At the longer time limit, mean-square displacement becomes independent of $t$. We can get the exact expression of that $t$ independent steady state value from Eq.(\ref{eqn:msdx_kappa}) and Eq.(\ref{y2t}) putting $t$ as large when all the exponential terms vanish. In the limit $t\rightarrow\infty$, both $\langle x^2(t)\rangle$ and $\langle y^2(t)\rangle$ relax to the same stationary value, 
\begin{equation}
\begin{split}
\langle x^2\rangle_{t\rightarrow\infty}=\langle y^2\rangle_{t\rightarrow\infty}=\frac{k_BT}{\kappa}+\frac{v_0^2}{2\kappa\Bar{\Gamma}(\kappa\Bar{\Gamma}+D_\theta)}+\mathcal{O}(e^{-\lambda t})
\end{split}
\label{steady}
\end{equation}
where $\lambda=\min(2\kappa\Bar{\Gamma}, 4D_\theta, \kappa\Bar{\Gamma}+D_\theta)$ gives the leading order time-scale of relaxation to the stationary value for anisotropic particle. On the contrary, the leading-order time-scale of relaxation to the stationary value for an isotropic particle in a two dimensional harmonic trap is given by $\lambda=\min(2\kappa\Bar{\Gamma}, \kappa\Bar{\Gamma}+D_\theta)$\cite{Chaudhuri_2021}. In general, we can conclude that the relaxation time scale is the same for both isotropic and anisotropic particles. However, for extremely small rotational diffusion constant $D_\theta$, when the particle rotates extremely slowly, the relaxation time scales differ in both cases. In such circumstances, the relaxation time scale for an anisotropic particle depends on $4D_\theta$. 

    \begin{figure}[h]
	    \centering
	     \includegraphics[width=0.9\textwidth]{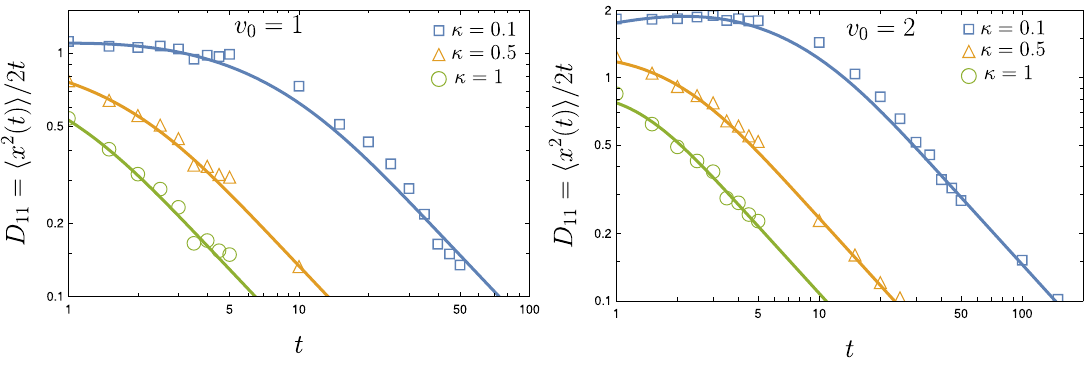}
	     \caption{Variation of diffusion coefficient along the $x$ axis with time in harmonic potential for different $v_0$ and $\kappa$ values as indicated in the figure. The translational diffusivities are fixed at $D_{\parallel}=1.0$, $D_{\perp}=0.9$. The rotational diffusivity and the initial angle $\theta_0$ are fixed at $D_\theta=1$ and $\theta_0=0$ respectively. Solid lines are the representations of analytical expression drawn from Eq.(\ref{eqn:msdx_kappa}) calculating $\langle x^2(t)\rangle/2t$.}
	    \label{fig:diff_kappa}
	    \end{figure}

Eq.(\ref{steady}) clearly shows the steady-state expression for the mean-square displacement of the particle in the harmonic potential. Here we can see that MSD is inversely proportional to the stiffness coefficient $\kappa$, which explains how drastically a steady state is attained for the stronger harmonic trap. Furthermore, at the longer time scale, MSD is also proportional to $v_0^2$. Results found for large time scale in Eq.(\ref{steady}) are shown in fig.(\ref{fig:msd_kappa1}) and fig.(\ref{fig:msd_kappa2}). Fig.(\ref{fig:diff_kappa}) shows the variation of diffusion coefficient with time in a harmonic potential, which shows that diffusive behavior sustains for longer time for higher values of propulsion velocity for any fixed value of stiffness coefficient $\kappa$. Let us now investigate the change in the relative diffusion coefficient over time for different velocities, while keeping other parameters fixed. If we expand the exponential terms of Eq.(\ref{eqn:msdx_kappa}) up to the first order of $t$ and compute the diffusion coefficient as $t\approx 0$, we find the expression, assuming $k_BT=1$(which implies $\Delta D=\Delta\Gamma$ and $\Bar{D}=\Bar{\Gamma}$), as follows:
\begin{equation}
D_{11}(t\approx 0)=D_{\parallel}+\frac{\kappa^2\Bar{D}\Delta D^2}{4(\kappa\Bar{\Gamma}+2D_\theta)(\kappa\Bar{\Gamma}+4D_\theta)}+\frac{3v_0^2D_\theta\kappa\Delta D}{4(\kappa\Bar{\Gamma}+2D_\theta)(\kappa\Bar{\Gamma}+D_\theta)(\kappa\Bar{\Gamma}-D_\theta)}
\label{relative}
\end{equation}  
 
    \begin{figure}[h]
	    \centering
	     \includegraphics[width=0.7\textwidth]{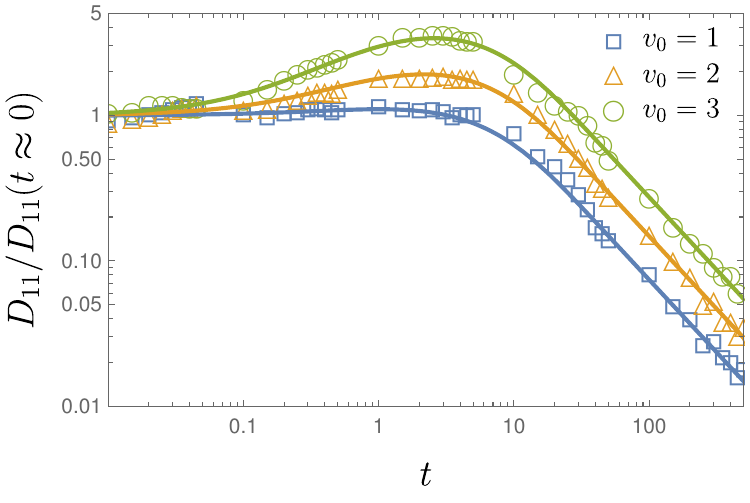}
	     \caption{Variation of relative diffusion coefficient along the $x$ axis with time in harmonic potential for different $v_0$ shown in the figure. The translational diffusivities are fixed at $D_{\parallel}=1.0$, $D_{\perp}=0.9$. The rotational diffusivity, the initial angle $\theta_0$ and $\kappa$ are fixed at $D_\theta=1$, $\theta_0=0$ and $\kappa=0.1$ respectively. Solid lines are the representations of analytical expression drawn from Eq.(\ref{eqn:msdx_kappa}) calculating $D_{11}=\langle x^2(t)\rangle/2t$ divided by $D_{11}(t\approx 0)$ of Eq.(\ref{relative}).}
	    \label{fig:diff_rel}
	    \end{figure}	
From fig.(\ref{fig:diff_rel}), we observe that the relative diffusion coefficient remains sustained for a longer duration at higher values of propulsion velocity $v_0$.

 \begin{figure}
  \centering
  \subfigure[Free particle]{\label{traj_free}\includegraphics[width=0.32\textwidth]{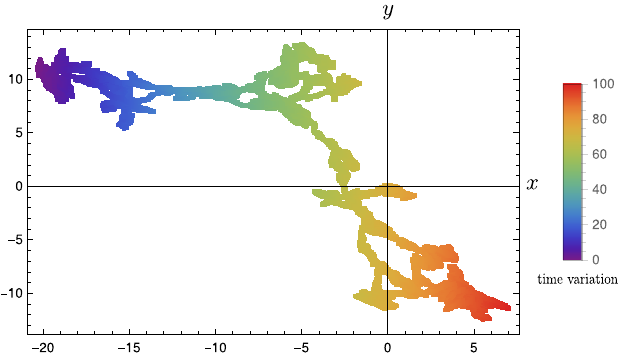}}
  \hfill
  \subfigure[Particle in a harmonic trap with $\kappa=0.1$]{\label{traj_k0.1}\includegraphics[width=0.32\textwidth]{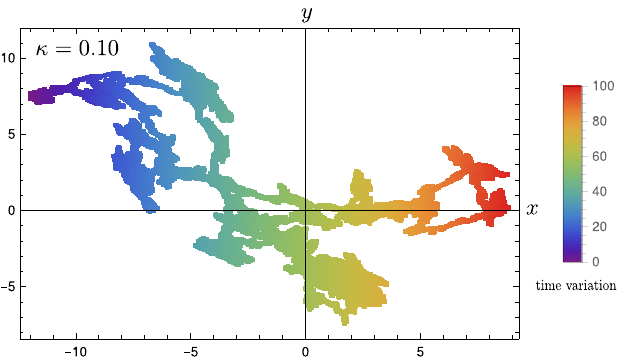}}
  \hfill
  \subfigure[Particle in a harmonic trap with $\kappa=1.0$]{\label{traj_k1.0}\includegraphics[width=0.32\textwidth]{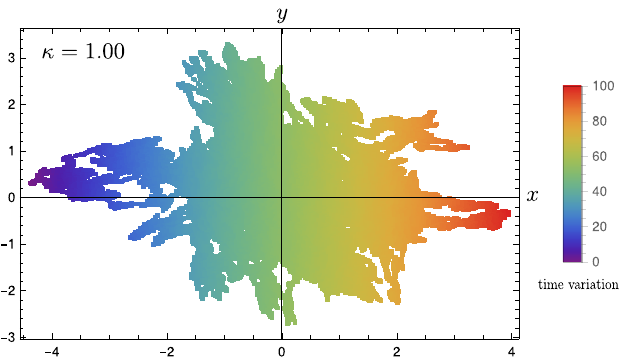}}
  \caption{Trajectory of the particle for $D_\parallel=1.0$, $D_\perp=0.1$ and $v_0=1.0$. The initial orientational angle and rotational diffusion coefficient are fixed at $\theta_0=0$ and $D_\theta=0.1$ respectively.}
  \label{trajectory}
\end{figure}
	
	 The trajectory of the particle is depicted in fig.(\ref{trajectory}), both in the absence of any potential and under the influence of harmonic confinement. The discernible trend is that the particle's motion progressively becomes constrained with an increase in the strength of the harmonic trap. This observation aligns with the findings of the mean-square displacement analysis. When there is no potential, the particle exhibits a transition of regimes, starting from an initially diffusive phase, evolving into a super-diffusive state, and ultimately returning to a diffusive behavior in the longer time duration, as illustrated in fig.(\ref{traj_free}). When we apply external harmonic potential, the motion becomes much more localized and a steady state is achieved depending on the strength of the stiffness coefficient as shown in figs.(\ref{traj_k0.1}) and (\ref{traj_k1.0}). For stronger confinement the steady state is achieved earlier. 
	 
\subsection{In a constant force field}

In this section, we compute the temporal evolution of the diffusion coefficients for an active ellipsoidal particle experiencing a constant force field ($F_x$ and $F_y$ represent the $x$ and $y$ components, respectively, of the constant force), while confined to a plane. An example would be an asymmetric particle uniformly charged and subjected to an electric field\cite{grima2007brownian}. We have determined the analytical expression of mean-square displacements for an active ellipsoidal particle in a constant force field using Eq.(\ref{lab_frame})(detailed calculations are in Appendix(\ref{const_force})). The expressions are written below

\begin{figure}[h]
	    \centering
	     \includegraphics[width=0.70\textwidth]{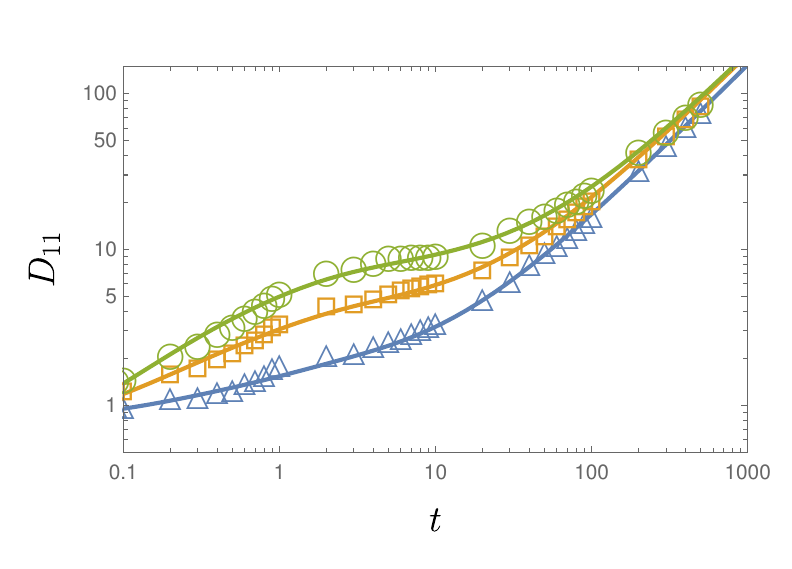}
	    \caption{ The variation of translational diffusion coefficient in the $x$ direction $D_{11}$ with time
$t$ for an ellipsoidal particle with parameters $D_\parallel =1$, $D_\perp = 0.1$, $D_\theta=1$, and $\theta_0=0$ in  the presence of constant external force $F_x=1$ and $F_y=0$ for different $v_0$ values; $v_0=1$ (Open blue triangles), $v_0=2$ (open orange squares), and $v_0=3$ (open green circles). Solid lines are to express the analytical result of Eq.(\ref{eqn:msd_conf}) calculating $\langle x^2(t)\rangle/2t$. }

	    \label{fig:cons_diff}
	    \end{figure}

\begin{equation}
    \begin{split}
    \langle x^2(t)\rangle&=F_x^2\Bar{\Gamma}^2t^2+v_0^2(\cos{2\theta_0}I_3+I_4)+\frac{\Delta\Gamma^2}{4}\Bigg[F_x^2(\cos{4\theta_0}I_a+I_b)+F_y^2(I_b-\cos{4\theta_0}I_a)\Bigg]\\
    &+2k_BT\Bar{\Gamma}t+k_BT\Delta\Gamma\cos{2\theta_0}\tau_4+2v_0F_x\Bar{\Gamma}t\cos{\theta_0}\tau_1+v_0F_x\Delta\Gamma\Big(\cos{\theta_0}I_2+\cos{3\theta_0}I_1\Big)\\
    &+v_0F_y\Delta\Gamma\Big(\sin{3\theta_0}I_1+\sin{\theta_0}I_2\Big)+F_x^2\Bar{\Gamma}\Delta\Gamma t\cos{2\theta_0}\tau_4+F_x\Bar{\Gamma}tF_y\Delta\Gamma\sin{2\theta_0}\tau_4+\frac{F_xF_y\Delta\Gamma^2}{2}\sin{4\theta_0}I_a
    \end{split}
    \label{eqn:msd_conf}
\end{equation}

\begin{equation}
    \begin{split}
    \langle y^2(t)\rangle&=F_y^2\Bar{\Gamma}^2t^2+v_0^2(I_4-\cos{2\theta_0}I_3)+\frac{\Delta\Gamma^2}{4}\Bigg[F_x^2(I_b-\cos{4\theta_0}I_a)+F_y^2(I_b+\cos{4\theta_0}I_a)\Bigg]\\
    &+2k_BT\Bar{\Gamma}t-k_BT\Delta\Gamma\cos{2\theta_0}\tau_4+2v_0F_y\Bar{\Gamma}t\sin{\theta_0}\tau_1-v_0F_y\Delta\Gamma\Big(\sin{\theta_0}I_2+\sin{3\theta_0}I_1\Big)\\
    &+v_0F_x\Delta\Gamma\Big(\cos{\theta_0}I_2-\cos{3\theta_0}I_1\Big)-F_y^2\Bar{\Gamma}\Delta\Gamma t\cos{2\theta_0}\tau_4+F_y\Bar{\Gamma}tF_x\Delta\Gamma\sin{2\theta_0}\tau_4-\frac{F_xF_y\Delta\Gamma^2}{2}\sin{4\theta_0}I_a
    \end{split}
    \label{MSDconsty}
\end{equation}	

where $\tau_n(t)$ is defined as $\tau_n(t)=\frac{1-e^{-nD_\theta t}}{nD_\theta}$. In fig.(\ref{fig:cons_diff}) we have shown the numerical results for diffusion coefficient along $x$ axis in a constant force field for different propulsion velocities. The numerical results are validating the analytical expression found in Eq.(\ref{eqn:msd_conf}) if we write as $D_{11}=\langle x^2(t)\rangle/2t$. Both analytical and simulation results indicate that the diffusion coefficient $D_{11}$ becomes equal for all propulsion velocities at the longer time. This shows that at longer times, an increment in propulsion velocity does not affect the diffusion coefficient; only the effect of the constant force remains dominant. 
     
\subsection{Torque and rotational diffusion in constant force field}	    
As demonstrated in the case of a constant external force along the $x$ axis, a finite torque is exerted on the particle as shown in the fig.(\ref{fig:torque}). Certainly, the strength of the torque is directly proportional to the strength of the applied force. The influence of this external torque manifests in the particle's rotational dynamics, rendering it super-diffusive during an intermediate time regime, as illustrated in fig.(\ref{fig:rot_diff}). However, in both the transient and long-time regimes, the rotational dynamics exhibit a purely diffusive nature, as depicted in the same figure. Variations in the strength of the constant force will impact the torque exerted on the particle, consequently influencing the super-diffusive nature observed in the rotational dynamics. Calculation of rotational MSD is shown in the Appendix(\ref{MSDrot}).  Indeed, the rotational dynamics of the particle is contingent upon both the nature and strength of the applied force. The type of force and its magnitude play crucial roles in determining the rotational behavior, with variations potentially leading to different rotational dynamics, including super-diffusive or purely diffusive characteristics, as illustrated in the context of our described scenarios.

  \begin{figure}
  \centering
  \subfigure[Variation of torque production with time $t$]{\label{fig:torque}\includegraphics[width=0.49\textwidth]{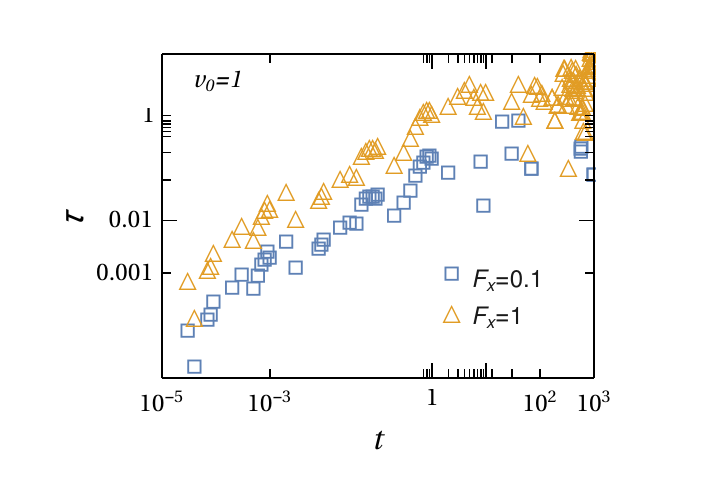}}
  \hfill
  \subfigure[The variation of rotational mean-square displacement with time $t$. Solid lines are analytical expression of Eq.(\ref{finaltaumsd}).]{\label{fig:rot_diff}\includegraphics[width=0.49\textwidth]{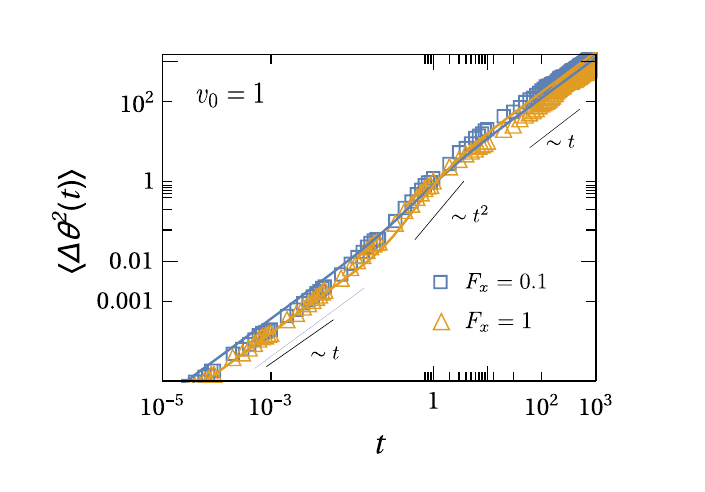}}
  \caption{The set of translational diffusivity $D_{\parallel}=1$ and $D_\perp=0.1$ and $D_\theta=1$ for different strengths of force shown in the figure, initial orientation angle is kept fixed at $\theta_0=0$. }
\end{figure}

	\subsection{Scaled average velocity}    
To facilitate our discourse, we introduce the concept of the scaled average velocity, denoted as $V_s = v/v_0$ in this section. In an active free ellipsoidal particle, the scaled average velocity refers to the average velocity of the particle normalized by its active velocity. This quantity is often used to understand how the motion of the particle changes with increasing active velocity. We calculated the average velocity along $x$ axis from Eq.(\ref{lab_frame})

\begin{equation}
v_{\theta_0}=\lim_{t \to \infty} \frac{\langle x(t) \rangle_{\theta_0}^{\eta_1,\eta_2}}{t}
\end{equation}

where $\theta_0$ is initial angle of orientation the particle. The full average
velocity after a second average over all $\theta_0$ is

\begin{equation}
v=\frac{1}{2\pi}\int_{0}^{2\pi} v_{\theta_0}d\theta_0
\end{equation}

In our simulations, the velocity calculations involved an integration step time $\Delta t=10^{-4}$, and the total integration time exceeded $3 \times 10^5$ to ensure accurate results. Additionally, we accounted for transient effects by estimating and subtracting their contributions. The stochastic averages presented above were computed through ensemble averages over $3\times 10^4$ trajectories, each characterized by random initial conditions. In our numerical simulations, the primary focus lies in the computation of the average velocity for two distinct scenarios: the particle subjected to a constant force and the particle experiencing a one-dimensional ratchet potential.

     \begin{figure}[h]
	    \centering
	     \includegraphics[width=0.7\textwidth]{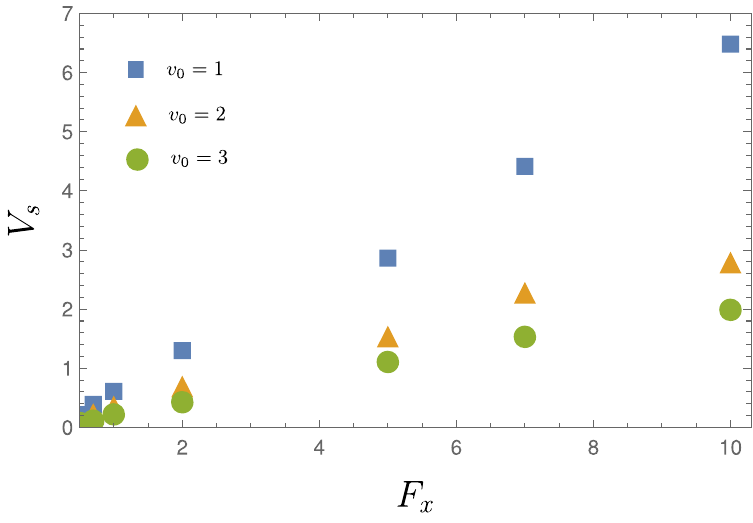}
	    \caption{Plot of scaled average velocity $V_s$ with constant force field ($F_x$) for different $v_0$ values as indicated in the legend. The $F_y$ component is maintained at 0. The translational diffusivities are set to $D_\parallel=1.0$ and $D_\perp=0.1$, while the rotational diffusivity is $D_\theta=1$.}
	    \label{fig:cons_vel}
	    \end{figure}	
 \begin{figure}
  \centering
  \subfigure[$V_s$ as a function of the asymmetry parameter $\Delta\Gamma$ for different $v_0$ values(indicated in the legend) and $D_\theta=1$.]{\label{VsGamma}\includegraphics[width=0.49\textwidth]{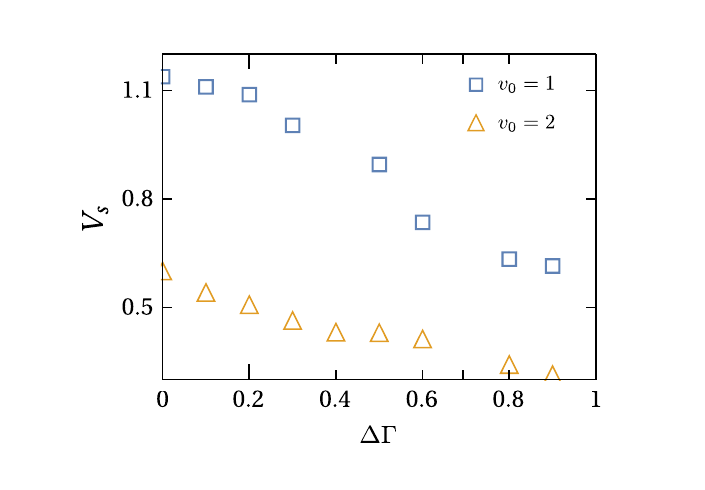}}
  \hfill
  \subfigure[$V_s$ as a function of $D_\theta$ for different $\Delta\Gamma$ values(indicated in the legend) and $v_0=1$]{\label{VsDtheta}\includegraphics[width=0.49\textwidth]{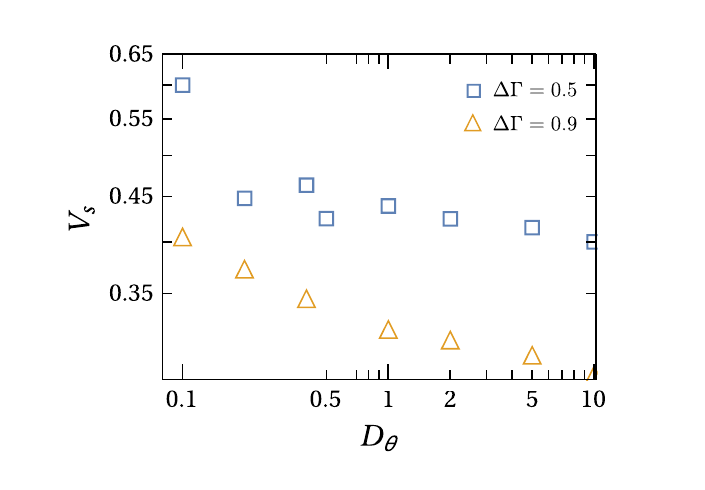}}
  \caption{Variation of Scaled average velocity in the constant force field $F_x=1$.}
\end{figure}
\subsubsection{In a constant force field}
Fig.(\ref{fig:cons_vel}) shows that increasing constant force along $x$ axis($F_x$) always pushes the particle towards increasing particular directed transport. Here the constant force is a very specifically directed force, so the force dominates to decide the directed transport. The relationship between the scaled average velocity $V_s$ and the asymmetrical parameter $\Delta\Gamma$ of the particle is depicted in the fig.(\ref{VsGamma}) for $D_\theta=1$. It is observed that the scaled average velocity $V_s$ exhibits a monotonous decrease with an increase in the asymmetrical parameter $\Delta\Gamma$. A higher asymmetric parameter indicates stronger forces opposing the preferred direction of the particle's motion, leading to a reduction in the scaled average velocity of the particle. The outcomes illustrating $V_s$ as a function of the rotational diffusion rate $D_\theta$ are showcased in fig(\ref{VsDtheta}), encompassing diverse values of the parameter $\Delta\Gamma$. The observation indicates that $V_s$ decreases with increasing values of $D_\theta$. In the limit as $D_\theta\rightarrow 0$, the self-propulsion angle $\theta(t)$ undergoes minimal changes, and the scaled average velocity approaches its maximum value. When the particle rotates more rapidly, the directional transport of the particle diminishes. Therefore, if $D_\theta$ increases, the particle is more prone to random changes in its orientation, making it challenging to maintain a consistent directed motion. The cumulative impact of heightened angular diffusion is that the particle devotes more time to random changes in direction rather than moving consistently in a preferred direction. This phenomenon leads to a decline in the overall average velocity over time.


\subsubsection{1D ratchet potential}    
In this section, we conducted a study on the dynamic behavior of the system subject to an one-dimensional asymmetric ratchet potential given by
\begin{equation}
 U(y,t)=V(t)[\sin({2\pi y/\lambda})+\alpha\sin({4\pi y/\lambda})]
 \label{potential}
\end{equation}
\begin{figure}[h]
	    \centering
	     \includegraphics[width=0.7\textwidth]{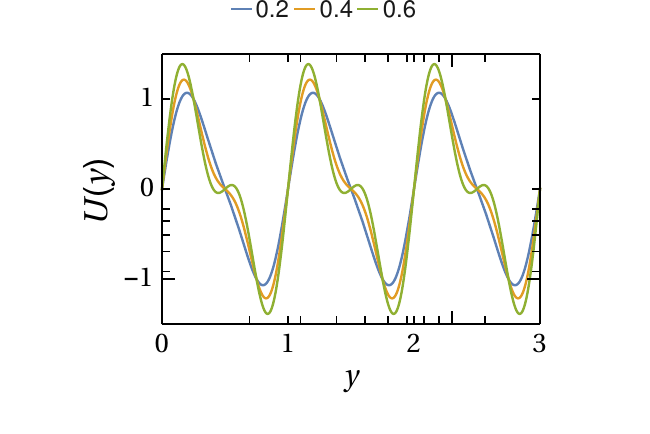}
	     \caption{Plot of the potential stated in Eq.(\ref{potential}) for different values of $\alpha$ indicated in the legend. }
	     \label{ratchet1}
	    \end{figure}
In the context of this study, $V(t)$ denotes the magnitude of the potential, where $V(t) = V(0) = 1.00$ in our specific case. The parameter $\alpha$ characterizes the degree of asymmetry in the potential, and we have chosen a periodicity of $\lambda = 1$ for the external potential. As the potential is oriented along the y-direction, we have quantified the scaled average velocity specifically in the y-direction during our investigation shown in the fig.(\ref{ratchet1}).
   \begin{figure}
  \centering
  \subfigure[$V_s$ as a function of the propulsion velocity $v_0$ for different $D_\theta$ values(indicated in the legend) and $\alpha=0.2$.]{\label{periodic_v0}\includegraphics[width=0.49\textwidth]{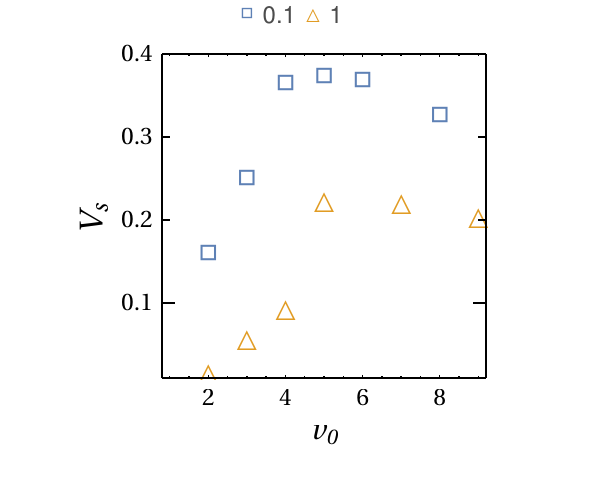}}
  \hfill
  \subfigure[$V_s$ as a function of the asymmetry parameter of the ratchet potential $\alpha$ for different $v_0$ values(indicated in the legend) and $D_\theta=0.1$]{\label{alpha}\includegraphics[width=0.49\textwidth]{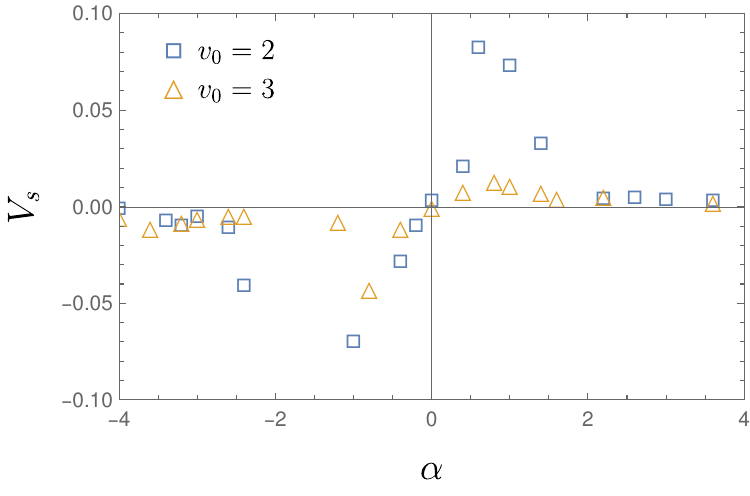}}
  \caption{Variation of Scaled average velocity in the ratchet potential with translational diffusion coefficients $D_\parallel=1.0$ and $D_\perp=0.1$.}
\end{figure}

Fig.(\ref{periodic_v0}) illustrates the scaled average velocity $V_s$ plotted against the self-propelled velocity $v_0$. As the parameter $v_0$ increases, there is a concomitant elevation in the input energy of the particle, resulting in an associated increase in its velocity. At higher values of $v_0$, the impact of the potential's asymmetry diminishes, leading to a reduction in the scaled velocity ($V_s$) of the particle. Consequently, an optimal parameter value $v_0$ arises, facilitating the attainment of the maximum scaled velocity $V_s$. Indeed, the identification of an optimal self-propelled velocity can facilitate the rectification of particles, enhancing their directed motion in the presence of an asymmetric potential. In fig.(\ref{alpha}), we have presented the variations of the scaled average velocity $V_s$ with the asymmetric parameter $\alpha$ of the potential. This behavior can be elucidated by referencing fig.(\ref{ratchet1}). The simulation for fig.(\ref{alpha}) is conducted over $10^6$ time steps, considering $12$ different initial angle configurations. The figure demonstrates that for positive $\alpha$, the scaled average velocity $V_s$ is positive, whereas for negative $\alpha$, it reverses. The particle can move in both left and right directions within a given time step, with this movement being determined by a time-averaged value. In the case of an active Brownian particle, an asymmetric potential alone can generate a net drift speed. The average velocity, which is scaled, depends on the asymmetric parameter $\alpha$. When $\alpha=0$, the potential becomes symmetric, resulting in an equal probability for the particle to move left or right. For $\alpha>0$, the probability of the particle moving towards the right is greater compared to moving left. Consequently, over a certain time average, we observe a positive scaled average velocity value. For lower positive values of $\alpha$, the particle follows a more pronounced and steeper trajectory; conversely, as $\alpha$ increases, the scaled average velocity of the particle decreases. Similarly, for $\alpha<0$, the converse is true: the probability of the particle moving towards the left is greater compared to the right, resulting in a negative scaled average velocity.

\section{Conclusion}\label{Conclusion}

This article explores the diffusion dynamics of an active ellipsoidal particles in a two-dimensional system under the assumption of the overdamped limit, neglecting the effects of inertia. We investigate the impact of shape asymmetry and propulsion velocity on the dynamics of the active particle. It is well-known that the behavior is anisotropic for short times but becomes isotropic for longer times \cite{grima2007brownian}. We studied the short- and long-time diffusion dynamics of an active ellipsoidal particle in both a free system and under various potential fields. The movement of an asymmetric particle can be described by the Langevin equation with an isotropic translational diffusion coefficient. We demonstrated the variation of mean square displacement and diffusion coefficient with time for the anisotropic case in both a free system and a harmonically trapped system. Furthermore, we discussed the influence of propulsion velocity, diffusion coefficient, and asymmetry parameters on the diffusion dynamics of the particle. In the context of anisotropic free Active Brownian Particles (ABP), the evolving dynamics can be divided into three distinct phases. The initial phase demonstrates purely diffusive behavior, characterized by its transient nature. The second phase exhibits super-diffusive characteristics, contingent upon the propulsion velocity. Lastly, the third phase reverts to diffusive dynamics. An interesting phenomenon arises with particles in a harmonic trap, where typically the relaxation time scale remains consistent regardless of whether the particle is isotropic or anisotropic. However, for extremely slow rotation rates (i.e., very small $D_\theta$), a distinct difference in relaxation time scale emerges between anisotropic and isotropic particles. We have also discussed the scaled average velocity of a free active ellipsoidal particle and the particle under various force fields, including a constant force field and a one-dimensional ratchet. An interesting phenomenon occurs when the particle is in a one-dimensional ratchet. With a gradual increase in propulsion velocity, the scaled average velocity initially experiences growth. However, beyond a certain optimal propulsion velocity, the directionality of the particle's movement begins to decline. The outcomes showcased in this study hold significant relevance across a diverse array of applications in active anisotropic systems, encompassing domains such as bacterial swimmers, motor proteins, and motile cells.

\section*{ACKNOWLEDGMENTS}
We acknowledge Dr. Dipanjan Chakraborty of IISER, Mohali,  Prof. Sanjib Sabhapandit of RRI, Bengaluru and Shubhendu Shekhar Khali of IIT Madras for their valuable discussions.

\section*{Conflict of Interest}
The authors have no conflicts to disclose.
\section*{Author Contributions}
\textbf{Sudipta Mandal}: Conceptualization (equal); \textbf{Anirban Ghosh}: Conceptualization (equal).

	\newpage    
\bibliographystyle{unsrt}
\bibliography{reference}	

\newpage   

\newpage 
	    
\onecolumngrid
\begin{center}
{\bf {\Large{Appendix}}}
\end{center}
\begin{appendix}
\section{Calculation of Mean-Square displacement along $y$ direction in a harmonic confinement }\label{App}

To find out the expression for the MSD along $y$ axis, we take $i=j=y$ in the expression of Eq.(\ref{25}). We now calculate the expression up to the first order correction which is expressed as
\begin{equation}
\langle y^2(t)\rangle_{\eta,\theta}=\langle y^2_0(t)\rangle_{\eta,\theta}-(\kappa\Delta\Gamma)\langle y_0(t)y_1(t)\rangle_{\eta,\theta}
\label{msdycorr}
\end{equation}
Now we will calculate the correlation terms separately. The first term, which is the zero order correction term, can be expressed from Eq.(\ref{24})
\begin{equation}
    \begin{split}
        \langle y^2_0(t)\rangle&=\int_{0}^{t}dt^\prime\int_{0}^{t}dt^{\prime\prime}e^{-\kappa\Bar{\Gamma}(t-t^\prime)}e^{-\kappa\Bar{\Gamma}(t-t^{\prime\prime})}\langle\xi(t^\prime)\xi(t^{\prime\prime})\rangle+\int_{0}^{t}dt^\prime\int_{0}^{t}dt^{\prime\prime}e^{-\kappa\Bar{\Gamma}(t-t^\prime)}e^{-\kappa\Bar{\Gamma}(t-t^{\prime\prime})}
        v_0^2\langle \sin{\theta(t^\prime)}\sin{\theta(t^{\prime\prime})}\rangle
    \end{split}
    \label{a12}
\end{equation}
There are two integrals, lets say $i_1$ and $i_2$ respectively. Lets calculate these two separately
\begin{equation}
    \begin{split}
        i_1&=\int_{0}^{t}dt^\prime\int_{0}^{t}dt^{\prime\prime}e^{-\kappa\Bar{\Gamma}(t-t^\prime)}e^{-\kappa\Bar{\Gamma}(t-t^{\prime\prime})}\langle\xi(t^\prime)\xi(t^{\prime\prime})\rangle\\
        &=\int_{0}^{t}dt^\prime\int_{0}^{t}dt^{\prime\prime}e^{-\kappa\Bar{\Gamma}(t-t^\prime)}e^{-\kappa\Bar{\Gamma}(t-t^{\prime\prime})}\Big[\Bar{\Gamma}\mathbf{1}+\frac{\Delta\Gamma}{2}\Bar{\Bar{\mathcal{R}}}(t^\prime)\Big]\delta(t^\prime-t^{\prime\prime})\\
        &=2k_BTe^{-2\kappa\Bar{\Gamma}t}\int_{0}^{t}dt^\prime e^{2\kappa\Bar{\Gamma}t^\prime}\Big[\Bar{\Gamma}\mathbf{1}+\frac{\Delta\Gamma}{2}\langle\Bar{\Bar{\mathcal{R}}}(t^\prime)\rangle\Big]\\
        &=\frac{k_BT}{\kappa}\mathbf{1}(1-e^{-2\kappa\Bar{\Gamma}t})+2k_BTe^{-2\kappa\Bar{\Gamma}t}\int_{0}^{t}dt^\prime e^{2\kappa\Bar{\Gamma}t^\prime}\frac{\Delta\Gamma}{2}\Bar{\Bar{\mathcal{R}}}(\theta_0)e^{-4D_\theta t^\prime}\\
        &=\frac{k_BT}{\kappa}\mathbf{1}(1-e^{-2\kappa\Bar{\Gamma}t})+\Delta D\Bar{\Bar{\mathcal{R}}}(\theta_0)e^{-2\kappa\Bar{\Gamma}t}\Big(\frac{e^{(2\kappa\Bar{\Gamma}-4D_\theta)}-1}{2\kappa\Bar{\Gamma}-4D_\theta}\Big)\\
        &=\frac{k_BT}{\kappa}(1-e^{-2\kappa\Bar{\Gamma}t})-\Delta D\cos{2\theta_0}\Big(\frac{e^{-4D_\theta t}-e^{-2\kappa\Bar{\Gamma}t}}{2\kappa\Bar{\Gamma}-4D_\theta}\Big)
    \end{split}
    \label{A13}
\end{equation}
\begin{equation}
    \begin{split}
        i_2&=v_0^2e^{-2\kappa\Bar{\Gamma}t}\int_{0}^{t}dt^\prime\int_{0}^{t}dt^{\prime\prime}e^{\kappa\Bar{\Gamma}(t^\prime+t^{\prime\prime})}\langle \sin{\theta(t^\prime)}\sin{\theta(t^{\prime\prime})}\rangle\\
        &=\frac{v_0^2}{2}e^{-2\kappa\Bar{\Gamma}t}\int_{0}^{t}dt^\prime\int_{0}^{t}dt^{\prime\prime}e^{\kappa\Bar{\Gamma}(t^\prime+t^{\prime\prime})}\Big[e^{-D_\theta[t^\prime+t^{\prime\prime}-2 min(t^\prime, t^{\prime\prime})]}-\cos{2\theta_0}e^{-D_\theta[t^\prime+t^{\prime\prime}+2 min(t^\prime, t^{\prime\prime})]}\Big]\\
        &=\frac{v_0^2}{2}e^{-2\kappa\Bar{\Gamma}t}\Big[\int_{0}^{t}dt^\prime\int_{0}^{t^\prime}dt^{\prime\prime}e^{\kappa\Bar{\Gamma}(t^\prime+t^{\prime\prime})}e^{-D_\theta(t^\prime-t^{\prime\prime})}+\int_{0}^{t}dt^\prime\int_{t^\prime}^{t}dt^{\prime\prime}e^{\kappa\Bar{\Gamma}(t^\prime+t^{\prime\prime})}e^{-D_\theta(t^{\prime\prime}-t^\prime)}\Big]\\
        &-\frac{v_0^2\cos{2\theta_0}}{2}e^{-2\kappa\Bar{\Gamma}t}\Big[\int_{0}^{t}dt^\prime\int_{0}^{t^\prime}dt^{\prime\prime}e^{\kappa\Bar{\Gamma}(t^\prime+t^{\prime\prime})}e^{-D_\theta(t^\prime+3t^{\prime\prime})}+\int_{0}^{t}dt^\prime\int_{t^\prime}^{t}e^{\kappa\Bar{\Gamma}(t^\prime+t^{\prime\prime})}e^{-D_\theta(3t^\prime+t^{\prime\prime})}\Big]\\
        &=\frac{v_0^2}{2}\Big[\frac{1-2e^{-(\kappa\Bar{\Gamma}+D_\theta)t}+e^{-2\kappa\Bar{\Gamma}t}}{(\kappa\Bar{\Gamma}-D_\theta)(\kappa\Bar{\Gamma}+D_\theta)}-\frac{D_\theta(1-e^{-2\kappa\Bar{\Gamma}t})}{\kappa\Bar{\Gamma}(\kappa\Bar{\Gamma}-D_\theta)(\kappa\Bar{\Gamma}+D_\theta)}\Big]\\
       & -\frac{v_0^2\cos{2\theta_0}}{2}\Big[\frac{2D_\theta(e^{-4D_\theta t}-e^{-2\kappa\Bar{\Gamma}t})}{(2\kappa\Bar{\Gamma}-4D_\theta)(\kappa\Bar{\Gamma}-3D_\theta)(\kappa\Bar{\Gamma}-D_\theta)}+\frac{e^{-4D_\theta t}-2e^{-(\kappa\Bar{\Gamma}+D_\theta)t}+e^{-2\kappa\Bar{\Gamma}t}}{(\kappa\Bar{\Gamma}-D_\theta)(\kappa\Bar{\Gamma}-3D_\theta)}\Big]
    \end{split}
    \label{A15}
\end{equation}
So Eq.(\ref{a12}) finally takes the form 
\begin{equation}
\begin{split}
\langle y^2_0(t)\rangle&=\frac{k_BT}{\kappa}(1-e^{-2\kappa\Bar{\Gamma}t})-\Delta D\cos{2\theta_0}\Big(\frac{e^{-4D_\theta t}-e^{-2\kappa\Bar{\Gamma}t}}{2\kappa\Bar{\Gamma}-4D_\theta}\Big)+\frac{v_0^2}{2}\Big[\frac{1-2e^{-(\kappa\Bar{\Gamma}+D_\theta)t}+e^{-2\kappa\Bar{\Gamma}t}}{(\kappa\Bar{\Gamma}-D_\theta)(\kappa\Bar{\Gamma}+D_\theta)}-\frac{D_\theta(1-e^{-2\kappa\Bar{\Gamma}t})}{\kappa\Bar{\Gamma}(\kappa\Bar{\Gamma}-D_\theta)(\kappa\Bar{\Gamma}+D_\theta)}\Big]\\
&-\frac{v_0^2\cos{2\theta_0}}{2}\Big[\frac{2D_\theta(e^{-4D_\theta t}-e^{-2\kappa\Bar{\Gamma}t})}{(2\kappa\Bar{\Gamma}-4D_\theta)(\kappa\Bar{\Gamma}-3D_\theta)(\kappa\Bar{\Gamma}-D_\theta)}+\frac{e^{-4D_\theta t}-2e^{-(\kappa\Bar{\Gamma}+D_\theta)t}+e^{-2\kappa\Bar{\Gamma}t}}{(\kappa\Bar{\Gamma}-D_\theta)(\kappa\Bar{\Gamma}-3D_\theta)}\Big]
\end{split}
\label{firstordery}
\end{equation}
Now let us start calculating the first order correction term $\langle y_0(t)y_1(t)\rangle_{\eta,\theta}$ of Eq.(\ref{msdycorr}). To find out this term we will start using the values of $\bold{R}_0$ and $\bold{R}_1$ from Eq.(\ref{24}) and we find two time correlation,

\begin{equation}
\begin{split}
&\langle R_{0,i}(t_1)R_{1,j}(t_2)\rangle_{\eta,\theta}\\
&=\Big\langle R_{0,i}(t_1)\int_{0}^{t_2}dt_2^{\prime}e^{-\kappa\Bar{\Gamma}(t_2-t_2^\prime)}\sum_{k}\mathcal{R}_{jk}(t_2^\prime)R_{0,k}(t_2^\prime)\Big\rangle\\
&=\Big\langle\int_{0}^{t_2}dt_2^{\prime}e^{-\kappa\Bar{\Gamma}(t_2-t_2^\prime)}\sum_{k}\mathcal{R}_{jk}(t_2^\prime)R_{0,i}(t_1)R_{0,k}(t_2^\prime)\Big\rangle\\
&=\Big\langle\int_{0}^{t_2}dt_2^{\prime}e^{-\kappa\Bar{\Gamma}(t_2-t_2^\prime)}\sum_{k}\mathcal{R}_{jk}(t_2^\prime)\int_{0}^{t_1}dt_1^{\prime}\int_{0}^{t_2^\prime}dt_2^{\prime\prime}e^{-\kappa\Bar{\Gamma}(t_1-t_1^\prime)}e^{-\kappa\Bar{\Gamma}(t_2^\prime-t_2^{\prime\prime})}[\eta_i(t_1^\prime)\eta_j(t_2^{\prime\prime})+v_0^2\hat{n}_i(t_1^\prime)\hat{n}_j(t_2^{\prime\prime})]\Big\rangle
\end{split}
\label{ycorr}
\end{equation}

We evaluate the two terms in the integral separately. The first term involves an average over the thermal fluctuations in the position and we denote it by $i_3$. The explicit calculation of $i_3$ becomes

\begin{equation}
\begin{split}
i_3&=\int_{0}^{t_2}dt_2^{\prime}e^{-\kappa\Bar{\Gamma}(t_2-t_2^\prime)}\Big\langle\sum_{k}\mathcal{R}_{jk}(t_2^\prime)\int_{0}^{t_1}dt_1^{\prime}\int_{0}^{t_2^\prime}dt_2^{\prime\prime}e^{-\kappa\Bar{\Gamma}(t_1-t_1^\prime)}e^{-\kappa\Bar{\Gamma}(t_2^\prime-t_2^{\prime\prime})}\eta_i(t_1^\prime)\eta_j(t_2^{\prime\prime})\Big\rangle\\
&=\int_{0}^{t_2}dt_2^{\prime}e^{-\kappa\Bar{\Gamma}(t_2-t_2^\prime)}\Big\langle\sum_{k}\mathcal{R}_{jk}(t_2^\prime)\Big[\frac{k_BT}{\kappa}\delta_{ki}(e^{-\kappa\Bar{\Gamma}(t_1-t_2^\prime)})+k_BT\Delta\Gamma e^{-\kappa\Bar{\Gamma}(t_1+t_2^\prime)}\int_{0}^{min(t_1,t_2^{\prime\prime})}dt^{\prime\prime}e^{2\kappa\Bar{\Gamma}t^{\prime\prime}}\mathcal{R}_{ki}(t^{\prime\prime})\Big]\Big\rangle\\
&=\frac{k_BT}{\kappa}e^{-\kappa\Bar{\Gamma}(t_1+t_2)}\int_{0}^{t_2}dt_2^\prime e^{\kappa\Bar{\Gamma}t_2^\prime}\Big\langle\mathcal{R}_{ji}(t_2^{\prime})\Big\rangle(e^{\kappa\Bar{\Gamma}t_2^\prime}-e^{-\kappa\Bar{\Gamma}t_2^\prime})\\
&+k_BT\Delta\Gamma e^{-\kappa\Bar{\Gamma}(t_1+t_2)}\int_{0}^{min(t_1,t_2^\prime)}dt^{\prime\prime}e^{2\kappa\Bar{\Gamma}t^{\prime\prime}}\sum_{k}\Big\langle\mathcal{R}_{jk}(t_2^{\prime})\mathcal{R}_{ki}(t^{\prime\prime})\Big\rangle
\end{split}
\end{equation}
In order to proceed further with the calculation, we consider the correlation function $\langle y_0(t_1)y_1(t_2)\rangle$ so that the above expression becomes
\begin{equation}
\begin{split}
i_3&=\frac{k_BT}{\kappa}e^{-\kappa\Bar{\Gamma}(t_1+t_2)}\mathcal{R}_{ji}(\theta_0)\int_{0}^{t_2}dt_2^\prime e^{-4D_\theta t_2^\prime}(e^{-4D_\theta t_2^\prime}-1)\\
&+k_BT\Delta\Gamma e^{-\kappa\Bar{\Gamma}(t_1+t_2)}\int_{0}^{t_2}dt_2^\prime\int_{0}^{t_2^\prime}dt^{\prime\prime}e^{2\kappa\Bar{\Gamma}t^{\prime\prime}}e^{-4D_\theta(t_2^\prime+t^{\prime\prime}-2min(t_2^\prime, t^{\prime\prime}))}\\
&=-\frac{k_BT}{\kappa}\cos2\theta_0e^{-\kappa\Bar{\Gamma}t_1}\Bigg(\frac{e^{(\kappa\Bar{\Gamma}-4D_\theta)t_2}-e^{-\kappa\Bar{\Gamma}t_2}}{2\kappa\Bar{\Gamma}-4D_\theta}-\frac{e^{-\kappa\Bar{\Gamma}t_2}-e^{-(\kappa\Bar{\Gamma}+4D_\theta)t_2}}{4D_\theta}\Bigg)\\
&+\Big(\frac{k_BT}{\kappa}\Big)\Big(\frac{\Delta\Gamma}{2\Bar{\Gamma}}\Big)e^{-\kappa\Bar{\Gamma}t_1}\Bigg[\frac{e^{\kappa\Bar{\Gamma}t_2}-e^{-\kappa\Bar{\Gamma}t_2}}{2\kappa\Bar{\Gamma}+4D_\theta}-\frac{2\kappa\Bar{\Gamma}}{4D_\theta}\frac{e^{-\kappa\Bar{\Gamma}t_2}-e^{-(\kappa\Bar{\Gamma}+4D_\theta)t_2}}{\kappa\Bar{\Gamma}+4D_\theta}\Bigg]
\end{split}
\end{equation}
The second term of Eq.(\ref{ycorr}) is denoted by $i_4$ and is given as for $\langle y_0(t_1)y_1(t_2)\rangle$

\begin{equation}
\begin{split}
i_4&=v_0^2\int_{0}^{t_2}dt_2^\prime e^{-\kappa\Bar{\Gamma}(t_2-t_2^\prime)}\Big\langle\sum\mathcal{R}_{jk}(t_2^\prime)\int_{0}^{t_1}dt_1^\prime\int_{0}^{t_2^\prime}dt_2^{\prime\prime}e^{-\kappa\Bar{\Gamma}(t_1-t_1^\prime)}e^{-\kappa\Bar{\Gamma}(t_2^\prime-t_2^{\prime\prime})}\sin{\theta}(t_1^\prime)\sin{\theta}(t_2^{\prime\prime})\Big\rangle\\
&=v_0^2\int_{0}^{t_2}dt_2^\prime e^{-\kappa\Bar{\Gamma}(t_2-t_2^\prime)}\int_{0}^{t_1}dt_1^\prime\int_{0}^{t_2^\prime}dt_2^{\prime\prime}e^{-\kappa\Bar{\Gamma}(t_1-t_1^\prime)}e^{-\kappa\Bar{\Gamma}(t_2^\prime-t_2^{\prime\prime})}\\
&\times[\langle\cos{2\theta}(t_2^\prime)\sin{\theta}(t_1^\prime)\sin{\theta}(t_2^{\prime\prime})\rangle
+\langle\sin{2\theta}(t_2^\prime)\sin{\theta}(t_1^\prime)\sin{\theta}(t_2^{\prime\prime})\rangle]
\end{split}
\label{54}
\end{equation}
We now use trigonometric identities
\begin{equation}
\begin{split}
&\cos{2\theta}(t_2^\prime)\sin{\theta}(t_1^\prime)\sin{\theta}(t_2^{\prime\prime})=\frac{1}{4}\Bigg[\cos{\Big(2\theta(t_2^\prime)+\theta(t_1^\prime)-\theta(t_2^{\prime\prime})\Big)}+\cos{\Big(2\theta(t_2^\prime)-\theta(t_1^\prime)+\theta(t_2^{\prime\prime})\Big)}\\
&-\cos{\Big(2\theta(t_2^\prime)+\theta(t_1^\prime)+\theta(t_2^{\prime\prime})\Big)}-\cos{\Big(2\theta(t_2^\prime)-\theta(t_1^\prime)-\theta(t_2^{\prime\prime})\Big)}\Bigg]\\
&\sin{2\theta}(t_2^\prime)\sin{\theta}(t_1^\prime)\sin{\theta}(t_2^{\prime\prime})=\frac{1}{4}\Bigg[\sin{\Big(2\theta(t_2^\prime)+\theta(t_1^\prime)-\theta(t_2^{\prime\prime})\Big)}+\sin{\Big(2\theta(t_2^\prime)-\theta(t_1^\prime)+\theta(t_2^{\prime\prime})\Big)}\\
&-\sin{\Big(2\theta(t_2^\prime)+\theta(t_1^\prime)+\theta(t_2^{\prime\prime})\Big)}-\sin{\Big(2\theta(t_2^\prime)-\theta(t_1^\prime)-\theta(t_2^{\prime\prime})\Big)}\Bigg]
\end{split}
\end{equation}
to rewrite the triple product of the trigonometric functions averaged over the rotational noise as
\begin{equation}
\begin{split}
&(a)\langle\cos{\Big(2\theta(t_2^\prime)+\theta(t_1^\prime)-\theta(t_2^{\prime\prime})\Big)}\rangle=\cos2\theta_0e^{-D_\theta(4t_2^\prime+t_1^\prime+t_2^{\prime\prime}+4min(t_2^\prime,t_1^\prime)-4min(t_2^\prime,t_2^{\prime\prime})-2min(t_1^\prime,t_2^{\prime\prime}))}\\
&(b)\langle\cos{\Big(2\theta(t_2^\prime)-\theta(t_1^\prime)+\theta(t_2^{\prime\prime})\Big)}\rangle=\cos2\theta_0e^{-D_\theta(4t_2^\prime+t_1^\prime+t_2^{\prime\prime}-4min(t_2^\prime,t_1^\prime)+4min(t_2^\prime,t_2^{\prime\prime})-2min(t_1^\prime,t_2^{\prime\prime}))}\\
&(c)\langle\cos{\Big(2\theta(t_2^\prime)+\theta(t_1^\prime)+\theta(t_2^{\prime\prime})\Big)}\rangle=\cos4\theta_0e^{-D_\theta(4t_2^\prime+t_1^\prime+t_2^{\prime\prime}+4min(t_2^\prime,t_1^\prime)+4min(t_2^\prime,t_2^{\prime\prime})+2min(t_1^\prime,t_2^{\prime\prime}))}\\
&(d)\langle\cos{\Big(2\theta(t_2^\prime)-\theta(t_1^\prime)-\theta(t_2^{\prime\prime})\Big)}\rangle=e^{-D_\theta(4t_2^\prime+t_1^\prime+t_2^{\prime\prime}-4min(t_2^\prime,t_1^\prime)-4min(t_2^\prime,t_2^{\prime\prime})+2min(t_1^\prime,t_2^{\prime\prime}))}\\
&(e)\langle\sin{\Big(2\theta(t_2^\prime)-\theta(t_1^\prime)-\theta(t_2^{\prime\prime})\Big)}\rangle=e^{-D_\theta(4t_2^\prime+t_1^\prime+t_2^{\prime\prime}-4min(t_2^\prime,t_1^\prime)-4min(t_2^\prime,t_2^{\prime\prime})+2min(t_1^\prime,t_2^{\prime\prime}))}\\
&(f)\langle\sin{\Big(2\theta(t_2^\prime)+\theta(t_1^\prime)+\theta(t_2^{\prime\prime})\Big)}\rangle=\sin4\theta_0e^{-D_\theta(4t_2^\prime+t_1^\prime+t_2^{\prime\prime}+4min(t_2^\prime,t_1^\prime)+4min(t_2^\prime,t_2^{\prime\prime})+2min(t_1^\prime,t_2^{\prime\prime}))}\\
&(g)\langle\sin{\Big(2\theta(t_2^\prime)-\theta(t_1^\prime)+\theta(t_2^{\prime\prime})\Big)}\rangle=\sin2\theta_0e^{-D_\theta(4t_2^\prime+t_1^\prime+t_2^{\prime\prime}-4min(t_2^\prime,t_1^\prime)+4min(t_2^\prime,t_2^{\prime\prime})-2min(t_1^\prime,t_2^{\prime\prime}))}\\
&(h)\langle\sin{\Big(2\theta(t_2^\prime)+\theta(t_1^\prime)-\theta(t_2^{\prime\prime})\Big)}\rangle=\sin2\theta_0e^{-D_\theta(4t_2^\prime+t_1^\prime+t_2^{\prime\prime}+4min(t_2^\prime,t_1^\prime)-4min(t_2^\prime,t_2^{\prime\prime})-2min(t_1^\prime,t_2^{\prime\prime}))}
\end{split}
\label{trigono}
\end{equation}
Note that in the above set of equations, the time dependence in the right-hand side of the set of equations (a)–(d) is
identical to that of the set of equations (e)–(h). Let us calculate the integral $i_4$ by using Eq.(\ref{trigono}); we start by taking the fourth term, (d), i.e., $\langle\cos{\Big(2\theta(t_2^\prime)-\theta(t_1^\prime)-\theta(t_2^{\prime\prime})\Big)}\rangle$, and calculate the following separately:
\begin{equation}
\frac{v_0^2e^{-\kappa\Bar{\Gamma}(t_1+t_2)}}{4}\int_{0}^{t_2}dt_2^\prime\int_{0}^{t_1}dt_1^\prime\int_{0}^{t_2^\prime}dt_2^{\prime\prime}e^{\kappa\Bar{\Gamma}t_1^\prime}e^{\kappa\Bar{\Gamma}t_2{^\prime\prime}}e^{-D_\theta(4t_2^\prime+t_1^\prime+t_2^{\prime\prime}-4min(t_2^\prime,t_1^\prime)-4min(t_2^\prime,t_2^{\prime\prime})+2min(t_1^\prime,t_2^{\prime\prime}))}
\end{equation}
In the above integral, always $t_2^\prime> t_2^{\prime\prime}$. In the first case, let us take $t_1^\prime>t_2^\prime$

case(1), $t_1^\prime>t_2^\prime$
\begin{equation}
\begin{split}
&\frac{v_0^2e^{-\kappa\Bar{\Gamma}(t_1+t_2)}}{4}\int_{0}^{t_2}dt_2^\prime\int_{t_2^\prime}^{t_1}dt_1^\prime\int_{0}^{t_2^\prime}dt_2^{\prime\prime}e^{\kappa\Bar{\Gamma}t_1^\prime}e^{\kappa\Bar{\Gamma}t_2{^\prime\prime}}e^{-4D_\theta t_2^\prime}e^{-D_\theta t_1^\prime}e^{-D_\theta t_2^{\prime\prime}}e^{4D_\theta t_2^\prime}e^{4D_\theta t_2^{\prime\prime}}e^{-2D_\theta t_2^{\prime\prime}}\\
&=\frac{v_0^2e^{-D_\theta t_1}(e^{D_\theta t_2}-e^{-\kappa\Bar{\Gamma}t_2})}{4(\kappa\Bar{\Gamma}+D_\theta)^2(\kappa\Bar{\Gamma}-D_\theta)}-\frac{v_0^2e^{-\kappa\Bar{\Gamma}t_1}(e^{\kappa\Bar{\Gamma}t_2}-e^{-\kappa\Bar{\Gamma}t_2})}{8\kappa\Bar{\Gamma}(\kappa\Bar{\Gamma}-D_\theta)(\kappa\Bar{\Gamma}+D_\theta)}-\frac{t_2v_0^2e^{-\kappa\Bar{\Gamma}t_2}e^{-D_\theta t_1}}{4(\kappa\Bar{\Gamma}+D_\theta)(\kappa\Bar{\Gamma}-D_\theta)}+\frac{v_0^2e^{-\kappa\Bar{\Gamma}t_1}(e^{-D_\theta t_2}-e^{-\kappa\Bar{\Gamma}t_2})}{4(\kappa\Bar{\Gamma}+D_\theta)(\kappa\Bar{\Gamma}-D_\theta)^2}
\end{split}
\label{case1}
\end{equation}
case(2), $t_1^\prime<t_2^\prime$
\begin{equation}
\begin{split}
&\frac{v_0^2e^{-\kappa\Bar{\Gamma}(t_1+t_2)}}{4}\Bigg[\int_{0}^{t_2}dt_2^\prime\int_{0}^{t_2^\prime}dt_1^\prime\int_{0}^{t_1^\prime}dt_2^{\prime\prime}e^{\kappa\Bar{\Gamma}t_1^\prime}e^{\kappa\Bar{\Gamma}t_2^{\prime\prime}}e^{-4D_\theta t_2^\prime}e^{3D_\theta t_1^\prime}e^{3D_\theta t_2^{\prime\prime}}e^{-2D_\theta t_2^{\prime\prime}}\\
&+\int_{0}^{t_2}dt_2^\prime\int_{0}^{t_2^\prime}dt_1^\prime\int_{t_1^\prime}^{t_2^\prime}dt_2^{\prime\prime}e^{\kappa\Bar{\Gamma}t_1^\prime}e^{\kappa\Bar{\Gamma}t_2^{\prime\prime}}e^{-4D_\theta t_2^\prime}e^{3D_\theta t_1^\prime}e^{3D_\theta t_2^{\prime\prime}}e^{-2D_\theta t_1^{\prime}}\Bigg]\\
&=\frac{v_0^2e^{-\kappa\Bar{\Gamma}t_1}}{8(\kappa\Bar{\Gamma}+D_\theta)(\kappa\Bar{\Gamma}+2D_\theta)}\Bigg[\frac{\sinh{\kappa\Bar{\Gamma}t_2}}{\kappa\Bar{\Gamma}}-\frac{e^{-\kappa\Bar{\Gamma}t_2}-e^{-(\kappa\Bar{\Gamma}+4D_\theta)t_2}}{4D_\theta}\Bigg]-\frac{v_0^2e^{-\kappa\Bar{\Gamma}t_1}}{4(\kappa\Bar{\Gamma}+D_\theta)(\kappa\Bar{\Gamma}+3D_\theta)}\Bigg[\frac{e^{-D_\theta t_2}-e^{-\kappa\Bar{\Gamma}t_2}}{\kappa\Bar{\Gamma}-D_\theta}\\
&-\frac{e^{-\kappa\Bar{\Gamma}t_2}-e^{-(\kappa\Bar{\Gamma}+4D_\theta)t_2}}{4D_\theta}\Bigg]+\frac{v_0^2e^{-\kappa\Bar{\Gamma}t_1}}{4(\kappa\Bar{\Gamma}+D_\theta)(\kappa\Bar{\Gamma}+3D_\theta)}\Bigg[\frac{\sinh{\kappa\Bar{\Gamma}t_2}}{\kappa\Bar{\Gamma}}-\frac{e^{-D_\theta t_2}-e^{-\kappa\Bar{\Gamma}t_2}}{\kappa\Bar{\Gamma}-D_\theta}\Bigg]\\
&-\frac{v_0^2e^{-\kappa\Bar{\Gamma}t_1}}{8(\kappa\Bar{\Gamma}+2D_\theta)(\kappa\Bar{\Gamma}+3D_\theta)}\Bigg[\frac{\sinh{\kappa\Bar{\Gamma}t_2}}{\kappa\Bar{\Gamma}}-\frac{e^{-\kappa\Bar{\Gamma}t_2}-e^{-(\kappa\Bar{\Gamma}+4D_\theta)t_2}}{4D_\theta}\Bigg]
\end{split}
\label{case2}
\end{equation}
We now add Eqs.(\ref{case1}) and (\ref{case2}) to find out the contribution due to (d) term of Eq.(\ref{trigono})
\begin{equation}
\begin{split}
i_d&=\frac{v_0^2e^{-D_\theta t_1}(e^{D_\theta t_2}-e^{-\kappa\Bar{\Gamma}t_2})}{4(\kappa\Bar{\Gamma}+D_\theta)^2(\kappa\Bar{\Gamma}-D_\theta)}+\frac{t_2v_0^2e^{-\kappa\Bar{\Gamma}t_2}e^{-D_\theta t_1}}{4(\kappa\Bar{\Gamma}-D_\theta)(\kappa\Bar{\Gamma}+D_\theta)}-\frac{3D_\theta v_0^2e^{-\kappa\Bar{\Gamma}t_1}\sinh{\kappa\Bar{\Gamma}t_2}}{4\kappa\Bar{\Gamma}(\kappa\Bar{\Gamma}+D_\theta)(\kappa\Bar{\Gamma}+2D_\theta)(\kappa\Bar{\Gamma}-D_\theta)}\\
&-\frac{v_0^2e^{-\kappa\Bar{\Gamma}t_1}(e^{-\kappa\Bar{\Gamma}t_2}-e^{-(\kappa\Bar{\Gamma}+4D_\theta)t_2})}{16(\kappa\Bar{\Gamma}+2D_\theta)(\kappa\Bar{\Gamma}+D_\theta)(\kappa\Bar{\Gamma}+3D_\theta)}-\frac{v_0^2e^{-\kappa\Bar{\Gamma}t_1}(\kappa\Bar{\Gamma}-5D_\theta)(e^{-D_\theta t_2}-e^{-\kappa\Bar{\Gamma}t_2})}{4(\kappa\Bar{\Gamma}+D_\theta)(\kappa\Bar{\Gamma}+3D_\theta)(\kappa\Bar{\Gamma}-D_\theta)^2}
\end{split}
\label{id}
\end{equation}
The other terms are calculated the way the first term was calculated to find the exact expression of Eq.(\ref{54}). The results of integrals due to terms (a)–(c) are as follows:
\begin{equation}
    \begin{split}
        i_a&=\frac{v_0^2e^{-\kappa\bar{\Gamma}t_1}}{8\kappa\bar{\Gamma}(\kappa\bar{\Gamma}+5D_\theta)}\Big[\frac{e^{(\kappa\bar{\Gamma}-4D_\theta)t_2}-e^{-\kappa\bar{\Gamma}t_2}}{2\kappa\bar{\Gamma}-4D_\theta}-\frac{e^{-\kappa\bar{\Gamma}t_2}-e^{-(\kappa\bar{\Gamma}+4D_\theta)t_2}}{4D_\theta}\Big]-\frac{v_0^2e^{-\kappa\bar{\Gamma}t_1}}{4(\kappa\bar{\Gamma}+5D_\theta)(\kappa\bar{\Gamma}-5D_\theta)}\Big[\frac{e^{-9D_\theta t_2}-e^{-\kappa\bar{\Gamma}t_2}}{\kappa\bar{\Gamma}-9D_\theta}\\
        &-\frac{e^{-\kappa\bar{\Gamma}t_2}-e^{-(\kappa\bar{\Gamma}+4D_\theta)t_2}}{4D_\theta}\Big]+\frac{v_0^2e^{-\kappa\bar{\Gamma}t_1}}{4(\kappa\bar{\Gamma}+3D_\theta)(\kappa\bar{\Gamma}-3D_\theta)}\Big[\frac{e^{(\kappa\bar{\Gamma}-4D_\theta)t_2}-e^{-\kappa\bar{\Gamma}t_2}}{2\kappa\bar{\Gamma}-4D_\theta}+\frac{e^{-\kappa\bar{\Gamma}t_2}-e^{-D_\theta t_2}}{\kappa\bar{\Gamma}-D_\theta}\Big]-\frac{v_0^2e^{-\kappa\bar{\Gamma}t_2}}{8\kappa\bar{\Gamma}(\kappa\bar{\Gamma}+3D_\theta)}\\
        &\Big[\frac{e^{(\kappa\bar{\Gamma}-4D_\theta)t_2}-e^{-\kappa\bar{\Gamma}t_2}}{2\kappa\bar{\Gamma}-4D_\theta}-\frac{e^{-\kappa\bar{\Gamma}t_2}-e^{-(\kappa\bar{\Gamma}+4D_\theta)t_2}}{4D_\theta}\Big]+\frac{v_0^2e^{-\kappa\bar{\Gamma}t_1}}{4(\kappa\bar{\Gamma}-D_\theta)(\kappa\bar{\Gamma}+5D_\theta)}\Big[\frac{e^{-9D_\theta t_2}-e^{-\kappa\bar{\Gamma}t_2}}{\kappa\bar{\Gamma}-9D_\theta}-\frac{e^{(\kappa\bar{\Gamma}-4D_\theta)t_2}-e^{-\kappa\bar{\Gamma}t_2}}{2\kappa\bar{\Gamma}-4D_\theta}\Big]\\
        &+\frac{v_0^2e^{-D_\theta t_1}}{4(\kappa\bar{\Gamma}-D_\theta)(\kappa\bar{\Gamma}+5D_\theta)}\Big[\frac{e^{-3D_\theta t_2}-e^{-\kappa\bar{\Gamma}t_2}}{\kappa\bar{\Gamma}-3D_\theta}-\frac{e^{-\kappa\bar{\Gamma}t_2}-e^{-(\kappa\bar{\Gamma}+8D_\theta)t_2}}{8D_\theta}\Big]
    \end{split}
    \label{ia}
\end{equation}
\begin{equation}
    \begin{split}
        i_b&=\frac{v_0^2e^{-D_\theta t_1}}{4(\kappa\bar{\Gamma}-D_\theta)(\kappa\bar{\Gamma}-3D_\theta)}\Big[\frac{e^{-3D_\theta t_2}-e^{-\kappa\bar{\Gamma}t_2}}{(\kappa\bar{\Gamma}-3D_\theta)}-t_2e^{-\kappa\bar{\Gamma}t_2}\Big]+\frac{v_0^2e^{-\kappa\bar{\Gamma}t_1}}{4(\kappa\bar{\Gamma}-D_\theta)(\kappa\bar{\Gamma}-3D_\theta)}\Big[\frac{e^{-D_\theta t_2}-e^{-\kappa\bar{\Gamma}t_2}}{\kappa\bar{\Gamma}-D_\theta}\\
        &-\frac{e^{(\kappa\bar{\Gamma}-4D_\theta)t_2}-e^{-\kappa\bar{\Gamma}t_2}}{\kappa\bar{\Gamma}-2D_\theta}\Big]+\frac{v_0^2e^{-\kappa\bar{\Gamma}t_1}}{4(\kappa\bar{\Gamma}+5D_\theta)(\kappa\bar{\Gamma}-5D_\theta)}\Big[\frac{e^{-4D_\theta t_2}-e^{-\kappa\bar{\Gamma}t_2}}{\kappa\bar{\Gamma}-4D_\theta}-\frac{e^{-9D_\theta t_2}-e^{-\kappa\bar{\Gamma}t_2}}{\kappa\bar{\Gamma}-9D_\theta}\Big]\\
        &+\frac{v_0^2e^{-\kappa\bar{\Gamma}t_1}}{16\kappa\bar{\Gamma}(\kappa\bar{\Gamma}-3D_\theta)}\Big[\frac{e^{(\kappa\bar{\Gamma}-4D_\theta)t_2}-e^{-\kappa\bar{\Gamma}t_2}}{\kappa\bar{\Gamma}-2D_\theta}-\frac{e^{-\kappa\bar{\Gamma}t_2}-e^{-(\kappa\bar{\Gamma}+4D_\theta)t_2}}{2D_\theta}\Big]+\frac{v_0^2e^{-\kappa\bar{\Gamma}t_1}}{4(\kappa\bar{\Gamma}-3D_\theta)(\kappa\bar{\Gamma}+3D_\theta)}\Big[\frac{e^{-\kappa\bar{\Gamma}t_2}-e^{-(\kappa\bar{\Gamma}+4D_\theta)t_2}}{4D_\theta}\\
        &-\frac{e^{-D_\theta t_2}-e^{-\kappa\bar{\Gamma}t_2}}{\kappa\bar{\Gamma}-D_\theta}\Big]
    \end{split}
    \label{ib}
\end{equation}
\begin{equation}
    \begin{split}
        i_c&=\frac{v_0^2e^{-D_\theta t_1}}{4(\kappa\bar{\Gamma}-D_\theta)(\kappa\bar{\Gamma}-7D_\theta)}\Big[\frac{e^{-15D_\theta t_2}-e^{-\kappa\bar{\Gamma}t_2}}{\kappa\bar{\Gamma}-15D_\theta}-\frac{e^{-\kappa\bar{\Gamma}t_2}-e^{-(\kappa\bar{\Gamma}+8D_\theta)t_2}}{8D_\theta}\Big]+\frac{v_0^2e^{-\kappa\bar{\Gamma}t_1}}{4(\kappa\bar{\Gamma}-D_\theta)(\kappa\bar{\Gamma}-7D_\theta)}\Big[\frac{e^{-9D_\theta t_2}-e^{-\kappa\bar{\Gamma}t_2}}{\kappa\bar{\Gamma}-9D_\theta}\\
        &-\frac{e^{(\kappa\bar{\Gamma}-16D_\theta)t_2}-e^{-\kappa\bar{\Gamma}t_2}}{2(\kappa\bar{\Gamma}-8D_\theta)}\Big]+\frac{v_0^2e^{-\kappa\bar{\Gamma}t_1}}{8(\kappa\bar{\Gamma}-7D_\theta)(\kappa\bar{\Gamma}-6D_\theta)}\Big[\frac{e^{(\kappa\bar{\Gamma}-16D_\theta)t_2}-e^{-\kappa\bar{\Gamma}t_2}}{2\kappa\bar{\Gamma}-16D_\theta}-\frac{e^{-\kappa\bar{\Gamma}t_2}-e^{-(\kappa\bar{\Gamma}+4D_\theta)t_2}}{4D_\theta}\Big]\\
        &+\frac{v_0^2e^{-\kappa\bar{\Gamma}t_1}}{4(\kappa\bar{\Gamma}-5D_\theta)(\kappa\bar{\Gamma}-7D_\theta)}\Big[\frac{e^{-\kappa\bar{\Gamma}t_2}-e^{-(\kappa\bar{\Gamma}+4D_\theta)t_2}}{4D_\theta}-\frac{e^{-9D_\theta t_2}-e^{-\kappa\bar{\Gamma}t_2}}{\kappa\bar{\Gamma}-9D_\theta}\Big]+\frac{v_0^2e^{-\kappa\bar{\Gamma}t_1}}{4(\kappa\bar{\Gamma}-5D_\theta)(\kappa\bar{\Gamma}-7D_\theta)}\Big[\frac{e^{(\kappa\bar{\Gamma}-16D_\theta)t_2}-e^{-\kappa\bar{\Gamma}t_2}}{2\kappa\bar{\Gamma}-16D_\theta}\\
        &-\frac{e^{-9D_\theta t_2}-e^{-\kappa\bar{\Gamma}t_2}}{\kappa\bar{\Gamma}-9D_\theta}\Big]+\frac{v_0^2e^{-\kappa\bar{\Gamma}t_1}}{8(\kappa\bar{\Gamma}-5D_\theta)(\kappa\bar{\Gamma}-6D_\theta)}\Big[\frac{e^{-\kappa\bar{\Gamma}t_2}-e^{-(\kappa\bar{\Gamma}+4D_\theta)t_2}}{4D_\theta}-\frac{e^{(\kappa\bar{\Gamma}-16D_\theta)t_2}-e^{-\kappa\bar{\Gamma}t_2}}{2\kappa\bar{\Gamma}-16D_\theta}\Big]
    \end{split}
     \label{ic}
\end{equation}
Integral value to due term (e) of Eq.(\ref{trigono}) is similar to term (d) calculated in Eq.(\ref{id}). Terms due to (f)–(h) of Eq.(\ref{trigono}) will be zero for the initial orientational angle $\theta_0=0$ as the contribution due to $\sin{\theta_0}$ will make the term zero. Now, for simplification, we take only the term associated with $\sinh{\kappa\Bar{\Gamma}t}$ Eq.(\ref{id}). Similarly, another term of $\sinh{\kappa\Bar{\Gamma}t}$ will arise due to the contribution of term (e) of Eq.(\ref{trigono}).

After adding the relevant terms, the final expression for the two-time correlation term becomes
\begin{equation}
\begin{split}
\langle y_0(t_1)y_1(t_2)\rangle&=-\frac{k_BT}{\kappa}\cos2\theta_0e^{-\kappa\Bar{\Gamma}t_1}\Bigg(\frac{e^{(\kappa\Bar{\Gamma}-4D_\theta)t_2}-e^{-\kappa\Bar{\Gamma}t_2}}{2\kappa\Bar{\Gamma}-4D_\theta}-\frac{e^{-\kappa\Bar{\Gamma}t_2}-e^{-(\kappa\Bar{\Gamma}+4D_\theta)t_2}}{4D_\theta}\Bigg)\\
&+\Big(\frac{k_BT}{\kappa}\Big)\Big(\frac{\Delta\Gamma}{2\Bar{\Gamma}}\Big)e^{-\kappa\Bar{\Gamma}t_1}\Bigg[\frac{e^{\kappa\Bar{\Gamma}t_2}-e^{-\kappa\Bar{\Gamma}t_2}}{2\kappa\Bar{\Gamma}+4D_\theta}-\frac{2\kappa\Bar{\Gamma}}{4D_\theta}\frac{e^{-\kappa\Bar{\Gamma}t_2}-e^{-(\kappa\Bar{\Gamma}+4D_\theta)t_2}}{\kappa\Bar{\Gamma}+4D_\theta}\Bigg]+\frac{3D_\theta v_0^2e^{-\kappa\Bar{\Gamma}t_1}\sinh{\kappa\Bar{\Gamma}t_2}}{2\kappa\Bar{\Gamma}(\kappa\Bar{\Gamma}+D_\theta)(\kappa\Bar{\Gamma}+2D_\theta)(\kappa\Bar{\Gamma}-D_\theta)}
\end{split}
    \label{y0y1}
\end{equation}
From the above equation, the first order correction in $\langle y^2(t)\rangle$ is given by
\begin{equation}
\begin{split}
\langle y_0(t)y_1(t)\rangle&=-\frac{k_BT}{\kappa}\cos2\theta_0e^{-\kappa\Bar{\Gamma}t}\Bigg(\frac{e^{(\kappa\Bar{\Gamma}-4D_\theta)t}-e^{-\kappa\Bar{\Gamma}t}}{2\kappa\Bar{\Gamma}-4D_\theta}-\frac{e^{-\kappa\Bar{\Gamma}t}-e^{-(\kappa\Bar{\Gamma}+4D_\theta)t}}{4D_\theta}\Bigg)\\
&+\Big(\frac{k_BT}{\kappa}\Big)\Big(\frac{\Delta\Gamma}{2\Bar{\Gamma}}\Big)e^{-\kappa\Bar{\Gamma}t}\Bigg[\frac{e^{\kappa\Bar{\Gamma}t}-e^{-\kappa\Bar{\Gamma}t}}{2\kappa\Bar{\Gamma}+4D_\theta}-\frac{2\kappa\Bar{\Gamma}}{4D_\theta}\frac{e^{-\kappa\Bar{\Gamma}t}-e^{-(\kappa\Bar{\Gamma}+4D_\theta)t}}{\kappa\Bar{\Gamma}+4D_\theta}\Bigg]+\frac{3D_\theta v_0^2e^{-\kappa\Bar{\Gamma}t}\sinh{\kappa\Bar{\Gamma}t}}{2\kappa\Bar{\Gamma}(\kappa\Bar{\Gamma}+D_\theta)(\kappa\Bar{\Gamma}+2D_\theta)(\kappa\Bar{\Gamma}-D_\theta)}
\end{split}
    \label{y0ty1t}
\end{equation}
To find $\langle y^2(t)\rangle$, we take Eq.(\ref{firstordery}) and Eq.(\ref{y0ty1t})
\begin{equation}
\begin{split}
\langle y^2(t)\rangle&=\langle y_0^2(t)\rangle-(\kappa\Delta\Gamma)\langle y_0(t)y_1(t)\rangle\\
&=\frac{k_BT}{\kappa}(1-e^{-2\kappa\Bar{\Gamma}t})-\frac{k_BT\Delta\Gamma}{4D_\theta}\Bigg[e^{-2\kappa\Bar{\Gamma}t}-e^{-(2\kappa\Bar{\Gamma}+4D_\theta)t}-(\kappa\Delta\Gamma)\frac{e^{-2\kappa\Bar{\Gamma}t}-e^{-(2\kappa\Bar{\Gamma}+4D_\theta)t}}{\kappa\Bar{\Gamma}+4D_\theta}\Bigg]\\
&+\frac{v_0^2}{2}\Bigg[\frac{1-2e^{-(\kappa\Bar{\Gamma}+D_\theta)t}+e^{-2\kappa\Bar{\Gamma}t}}{(\kappa\Bar{\Gamma}-D_\theta)(\kappa\Bar{\Gamma}+D_\theta)}-\frac{D_\theta(1-e^{-2\kappa\Bar{\Gamma}t})}{\kappa\Bar{\Gamma}(\kappa\Bar{\Gamma}-D_\theta)(\kappa\Bar{\Gamma}+D_\theta)}\Bigg]-\frac{v_0^2\cos{2\theta_0}}{2}\Bigg[\frac{2D_\theta(e^{-4D_\theta t}-e^{-2\kappa\Bar{\Gamma}t})}{(2\kappa\Bar{\Gamma}-4D_\theta)(\kappa\Bar{\Gamma}-3D_\theta)(\kappa\Bar{\Gamma}-D_\theta)}\\
&+\frac{e^{-4D_\theta t}-2e^{-(\kappa\Bar{\Gamma}+D_\theta)t}+e^{-2\kappa\Bar{\Gamma}t}}{(\kappa\Bar{\Gamma}-D_\theta)(\kappa\Bar{\Gamma}-3D_\theta)}\Bigg]-(\kappa\Delta\Gamma)\sinh{\kappa\Bar{\Gamma}t}e^{-\kappa\Bar{\Gamma}t}\Bigg[\frac{3D_\theta v_0^2}{2\kappa\Bar{\Gamma}(\kappa\Bar{\Gamma}+D_\theta)(\kappa\Bar{\Gamma}-D_\theta)(\kappa\Bar{\Gamma}+2D_\theta)}+\frac{k_BT\Delta\Gamma}{2\kappa\Bar{\Gamma}(\kappa\Bar{\Gamma}+D_\theta)}\Bigg]
\end{split}
\label{y2t}
\end{equation}

\section{Calculation of MSD in Constant force field}\label{const_force}
From Eq.(\ref{lab_frame}) we can write,
\begin{equation}
\begin{split}
   & x(t)=v_0\int_{0}^{t}\cos{\theta(t^\prime)}dt^\prime+F_x\Bar{\Gamma}t+\frac{F_x\Delta\Gamma}{2}\int_{0}^{t}\cos{2\theta(t^\prime)}dt^\prime+\frac{F_y\Delta\Gamma}{2}\int_{0}^{t}\sin{2\theta(t^\prime)}dt^\prime+\int_{0}^{t}\eta_1(t^\prime)dt^\prime\\
   &y(t)=v_0\int_{0}^{t}\sin{\theta(t^\prime)}dt^\prime+F_y\Bar{\Gamma}t-\frac{F_y\Delta\Gamma}{2}\int_{0}^{t}\cos{2\theta(t^\prime)}dt^\prime+\frac{F_y\Delta\Gamma}{2}\int_{0}^{t}\sin{2\theta(t^\prime)}dt^\prime+\int_{0}^{t}\eta_2(t^\prime)dt^\prime
    \end{split}
    \label{11}
\end{equation}
From Eq.(\ref{11}) we can write the second moments as follows
\begin{equation}
    \begin{split}
        \langle x^2(t)\rangle&=v_0^2\int_{0}^{t}dt^\prime\int_{0}^{t}dt^{\prime\prime}\langle\cos{\theta(t^\prime)}\cos{\theta(t^{\prime\prime})}\rangle+F_x^2\Bar{\Gamma}^2t^2+\frac{\Delta\Gamma^2}{4}\int_{0}^{t}dt^\prime\int_{0}^{t}dt^{\prime\prime}\Big[F_x^2\langle\cos{2\theta(t^\prime)}\cos{2\theta(t^{\prime\prime})}\rangle\\
        &+F_y^2\langle\sin{2\theta(t^\prime)}\sin{2\theta(t^{\prime\prime})}\rangle\Big]+\int_{0}^{t}dt^\prime\int_{0}^{t}dt^{\prime\prime}\langle\eta_1(t^\prime)\eta_1(t^{\prime\prime})\rangle+2v_0F_xt\Bar{\Gamma}\int_{0}^{t}\langle \cos{\theta(t^\prime)}\rangle dt^\prime\\
        &+v_0F_x\Delta\Gamma\int_{0}^{t}dt^\prime\int_{0}^{t}dt^{\prime\prime}\langle\cos{\theta(t^\prime)}\cos{2\theta(t^{\prime\prime})}\rangle+v_0F_y\Delta\Gamma\int_{0}^{t}dt^\prime\int_{0}^{t}dt^{\prime\prime}\langle\cos{\theta(t^\prime)}\sin{2\theta(t^{\prime\prime})}\rangle\\
        &+F_x^2\Bar{\Gamma}\Delta\Gamma t\int_{0}^{t}\langle \cos{2\theta(t^\prime)}\rangle dt^\prime+F_xF_y\Bar{\Gamma}\Delta\Gamma t\int_{0}^{t}\langle\sin{2\theta(t^\prime)}\rangle dt^\prime+\frac{F_xF_y\Delta\Gamma^2}{2}\int_{0}^{t}dt^\prime\int_{0}^{t}dt^{\prime\prime}\langle\cos{2\theta(t^\prime)}\sin{2\theta(t^{\prime\prime})}\rangle
    \end{split}
    \label{constforcemsd}
\end{equation}

\begin{equation}
    \begin{split}
    \langle y^2(t)\rangle&=v_0^2\int_{0}^{t}dt^\prime\int_{0}^{t}dt^{\prime\prime}\langle\sin{\theta(t^\prime)}\sin{\theta(t^{\prime\prime})}\rangle+F_y^2\Bar{\Gamma}^2t^2+\frac{\Delta\Gamma^2}{4}\int_{0}^{t}dt^\prime\int_{0}^{t}dt^{\prime\prime}\Big[F_y^2\langle\cos{2\theta(t^\prime)}\cos{2\theta(t^{\prime\prime})}\rangle\\
        &+F_x^2\langle\sin{2\theta(t^\prime)}\sin{2\theta(t^{\prime\prime})}\rangle\Big]+\int_{0}^{t}dt^\prime\int_{0}^{t}dt^{\prime\prime}\langle\eta_2(t^\prime)\eta_2(t^{\prime\prime})\rangle+2v_0F_yt\Bar{\Gamma}\int_{0}^{t}\langle \sin{\theta(t^\prime)}\rangle dt^\prime\\
        &-v_0F_y\Delta\Gamma\int_{0}^{t}dt^\prime\int_{0}^{t}dt^{\prime\prime}\langle\sin{\theta(t^\prime)}\cos{2\theta(t^{\prime\prime})}\rangle+v_0F_x\Delta\Gamma\int_{0}^{t}dt^\prime\int_{0}^{t}dt^{\prime\prime}\langle\sin{\theta(t^\prime)}\sin{2\theta(t^{\prime\prime})}\rangle\\
        &-F_y^2\Bar{\Gamma}\Delta\Gamma t\int_{0}^{t}\langle \cos{2\theta(t^\prime)}\rangle dt^\prime+F_xF_y\Bar{\Gamma}\Delta\Gamma t\int_{0}^{t}\langle\sin{2\theta(t^\prime)}\rangle dt^\prime-\frac{F_xF_y\Delta\Gamma^2}{2}\int_{0}^{t}dt^\prime\int_{0}^{t}dt^{\prime\prime}\langle\cos{2\theta(t^\prime)}\sin{2\theta(t^{\prime\prime})}\rangle
    \end{split}
    \label{constforcemsdy}
\end{equation}
The terms involving $\langle \cos{\theta(t^\prime)}\eta(t^{\prime\prime})\rangle$, $\langle \sin{2\theta(t^\prime)}\eta(t^{\prime\prime})\rangle$, $\langle \cos{2\theta(t^\prime)}\eta(t^{\prime\prime})\rangle$ and $t\langle\eta(t^\prime)\rangle$ will contribute zero as $\eta$ is a zero average noise. 
 Each term in Eq.(\ref{constforcemsd}) needs to be computed separately by individually calculating the associated components. 
The first term of Eq.(\ref{constforcemsd}) consists of $\langle\cos{\theta(t^\prime)}\cos{\theta(t^{\prime\prime})}\rangle$ is calculated as
\begin{equation}
    \begin{split}
        &\int_{0}^{t}dt^\prime\int_{0}^{t}dt^{\prime\prime}\langle\cos{\theta(t^\prime)}\cos{\theta(t^{\prime\prime})}\rangle\\
        &=\frac{1}{2}\int_{0}^{t}dt^\prime\int_{0}^{t}dt^{\prime\prime}\Bigg[\Big\langle\cos{(2\theta_0+\Delta\theta(t^\prime)+\Delta\theta(t^{\prime\prime}))}\Big\rangle+\Big\langle\cos{(\Delta\theta(t^\prime)-\Delta\theta(t^{\prime\prime}))}\Big\rangle\Bigg]\\
        &=\frac{1}{2}\int_{0}^{t}dt^\prime\int_{0}^{t}dt^{\prime\prime}\Bigg[\cos{2\theta_0}e^{-D_\theta[t^\prime+t^{\prime\prime}+2\min(t^\prime,t^{\prime\prime})]}+e^{-D_\theta[t^\prime+t^{\prime\prime}-2\min(t^\prime,t^{\prime\prime})]}\Bigg]\\
        &=\cos{2\theta_0}I_3+I_4
    \end{split}
\end{equation}
If we calculate $I_3$ and $I_4$ separately we find that
\begin{equation}
    \begin{split}
    &I_3=\frac{1}{2}\int_{0}^{t}dt^\prime\int_{0}^{t}dt^{\prime\prime}e^{-D_\theta[t^\prime+t^{\prime\prime}+2\min(t^\prime,t^{\prime\prime})]}\\
    &=\frac{1}{2}\int_{0}^{t}dt^\prime\int_{0}^{t^\prime}dt^{\prime\prime}e^{-D_\theta(3t^{\prime\prime}+t^\prime)}+\frac{1}{2}\int_{0}^{t}dt^{\prime\prime}\int_{0}^{t^{\prime\prime}}dt^{\prime}e^{-D_\theta(3t^{\prime}+t^{\prime\prime})}\\
    &=\frac{1}{12D_\theta^2}\Big(3+e^{-4D_\theta t}-4e^{-D_\theta t}\Big)\\
    \end{split}
    \label{I34}
\end{equation}
In the similar way we can calculate $I_4=\frac{1}{D_\theta^2}\Big(e^{-D_\theta t}+D_\theta t-1\Big)$.

We have used the identity 
\begin{equation}
    \Big\langle e^{i[m\Delta \theta(t^\prime)\pm n\Delta \theta(t^{\prime\prime})]}\Big\rangle_{\theta_0}^{\eta}=e^{-D_\theta[m^2t^\prime+n^2t^{\prime\prime}\pm 2mn\min(t^\prime,t^{\prime\prime})]}
\end{equation}
Accordingly, the averages of the trigonometric functions over the
rotational noise, which are used in the calculations, take the form
\begin{equation}
    \begin{split}
        &\langle\cos{[\theta(t^\prime)-\theta(t^{\prime\prime})]}\rangle=e^{-D_\theta[t^\prime+t^{\prime\prime}-2\min(t^\prime,t^{\prime\prime})]}\\
        &\langle\cos{[\theta(t^\prime)+\theta(t^{\prime\prime})]}\rangle=\cos{2\theta_0}e^{-D_\theta[t^\prime+t^{\prime\prime}+2\min(t^\prime,t^{\prime\prime})]}\\
        &\langle\sin{[\theta(t^\prime)+\theta(t^{\prime\prime})]}\rangle=\sin{2\theta_0}e^{-D_\theta[t^\prime+t^{\prime\prime}+2\min(t^\prime,t^{\prime\prime})]}\\
        &\langle\sin{[\theta(t^\prime)-\theta(t^{\prime\prime})]}\rangle=0
    \end{split}
\end{equation}

Similarly to calculate the term consisting $\langle\cos{2\theta(t^\prime)}\cos{2\theta(t^{\prime\prime})}\rangle$ in Eq.(\ref{constforcemsd}) we approach as follows

\begin{equation}
    \begin{split}
        &\int_{0}^{t}dt^\prime\int_{0}^{t}dt^{\prime\prime}\langle\cos{2\theta(t^\prime)}\cos{2\theta(t^{\prime\prime})}\rangle\\
        &=\frac{1}{2}\int_{0}^{t}dt^\prime\int_{0}^{t}dt^{\prime\prime}\Bigg[\Big\langle\cos{(4\theta_0+2\Delta\theta(t^\prime+2\Delta\theta(t^{\prime\prime}))}\Big\rangle+\Big\langle\cos{(2\Delta\theta(t^\prime)-2\Delta\theta(t^{\prime\prime}))}\Big\rangle\Bigg]\\
        &=\frac{1}{2}\int_{0}^{t}dt^\prime\int_{0}^{t}dt^{\prime\prime}\Bigg[\cos{4\theta_0}e^{-4D_\theta[t^\prime+t^{\prime\prime}+2\min(t^\prime,t^{\prime\prime})]}+e^{-4D_\theta[t^\prime+t^{\prime\prime}-2\min(t^\prime,t^{\prime\prime})]}\Bigg]\\
        &=\cos{4\theta_0}I_a+I_b
    \end{split}
\end{equation}

where,

\begin{equation}
    \begin{split}
        I_a&=\frac{1}{2}\int_{0}^{t}dt^\prime\int_{0}^{t}dt^{\prime\prime}e^{-4D_\theta[t^\prime+t^{\prime\prime}+2\min(t^\prime,t^{\prime\prime})]}\\
        &=\frac{1}{2}\int_{0}^{t}dt^\prime\int_{0}^{t^\prime}dt^{\prime\prime}e^{-4D_\theta(3t^{\prime\prime}+t^\prime)}+\frac{1}{2}\int_{0}^{t}dt^{\prime\prime}\int_{0}^{t^{\prime\prime}}dt^{\prime}e^{-4D_\theta(t^{\prime\prime}+3t^\prime)}\\
        &=\frac{1}{192D_\theta^2}\Big(3+e^{-16D_\theta t}-4e^{-4D_\theta t}\Big)
    \end{split}
\end{equation}
and 
$I_b=\frac{1}{16D_\theta^2}\Big(4D_\theta t+e^{-4D_\theta t}-1\Big)$

Likewise, the remaining terms in Eq.(\ref{constforcemsd}) and (\ref{constforcemsdy}) can be determined by computing them individually, and the results are as follows:
\begin{equation}
\int_{0}^{t}\int_{0}^{t}\langle\sin{2\theta(t^\prime)}\sin{2\theta(t^{\prime\prime})}\rangle dt^\prime dt^{\prime\prime}=I_b-\cos{4\theta_0}I_a
\end{equation}
\begin{equation}
\int_{0}^{t}\int_{0}^{t}\langle\sin{2\theta(t^\prime)}\cos{2\theta(t^{\prime\prime})}\rangle dt^\prime dt^{\prime\prime}=\sin{4\theta_0}I_a
\end{equation}
\begin{equation}
\int_{0}^{t}\int_{0}^{t}\langle\cos{\theta(t^\prime)}\cos{2\theta(t^{\prime\prime})}\rangle dt^\prime dt^{\prime\prime}=\cos{3\theta_0}I_1+\cos{\theta_0}I_2
\end{equation}
\begin{equation}
\int_{0}^{t}\int_{0}^{t}\langle\cos{\theta(t^\prime)}\sin{2\theta(t^{\prime\prime})}\rangle dt^\prime dt^{\prime\prime}=\sin{3\theta_0}I_1+\sin{\theta_0}I_2
\end{equation}
\begin{equation}
\int_{0}^{t}\int_{0}^{t}\langle\cos{\theta(t^\prime)}\cos{\theta(t^{\prime\prime})}\rangle dt^\prime dt^{\prime\prime}=\cos{2\theta_0}I_3+I_4
\end{equation}
\begin{equation}
\int_{0}^{t}\int_{0}^{t}\langle\sin{\theta(t^\prime)}\sin{\theta(t^{\prime\prime})}\rangle dt^\prime dt^{\prime\prime}=I_4-\cos{2\theta_0}I_3
\end{equation}

\begin{equation}
\int_{0}^{t}dt^\prime\int_{0}^{t}dt^{\prime\prime}\langle\sin{\theta(t^\prime)}\cos{2\theta(t^{\prime\prime})}\rangle=\sin{\theta_0}I_2+\sin{3\theta_0}I_1
\end{equation}
\begin{equation}
\int_{0}^{t}dt^\prime\int_{0}^{t}dt^{\prime\prime}\langle\sin{\theta(t^\prime)}\sin{2\theta(t^{\prime\prime})}\rangle=\cos{\theta_0}I_2-\cos{3\theta_0}I_1
\end{equation}

when
\begin{equation}
    \begin{split}
        &I_1=\frac{1}{144D_\theta^2}\Big(8+e^{-9D_\theta t}-9e^{-D_\theta t}\Big)+\frac{1}{360D_\theta^2}\Big(5+4e^{-9D_\theta t}-9e^{-4D_\theta t}\Big)\\
        &I_2=\frac{1}{2D_\theta}\Big(e^{-D_\theta t}-1\Big)+\frac{1}{24D_\theta^2}\Big(3+e^{-4D_\theta t}-4e^{-D_\theta t}\Big)\\
        &I_3=\frac{1}{12D_\theta^2}\Big(3+e^{-4D_\theta t}-4e^{-D_\theta t}\Big)\\
        &I_4=\frac{1}{D_\theta^2}\Big(e^{-D_\theta t}+D_\theta t-1\Big)\\
        &I_a=\frac{1}{192D_\theta^2}\Big(3+e^{-16D_\theta t}-4e^{-4D_\theta t}\Big)\\
        &I_b=\frac{1}{16D_\theta^2}\Big(4D_\theta t+e^{-4D_\theta t}-1\Big)
    \end{split}
\end{equation}
\subsection{Rotational MSD calculation}\label{MSDrot}
From Eq.(\ref{lab_frame}) we get the Langevin equation for the rotational motion, that gives
\begin{equation}
    \theta(t)=\int_{0}^{t}\Big[\eta_3(t^\prime)+\Gamma_\theta\tau(t^\prime)\Big]dt^\prime+\theta_0
\end{equation}
where $\theta_0$ is the initial orientational angle which is taken as $\theta_0=0$. So rotational MSD turns out to be
\begin{equation}
    \langle\Delta\theta^2(t)\rangle=\int_{0}^{t}dt^\prime\int_{0}^{t}dt^{\prime\prime}\langle\eta_3(t^\prime)\eta_3(t^{\prime\prime})\rangle+\Gamma^2_\theta\int_{0}^{t}dt^\prime\int_{0}^{t}dt^{\prime\prime}\langle\tau(t^\prime)\tau(t^{\prime\prime})\rangle
    \label{taumsd}
\end{equation}
The first integral comes from the fluctuation, that can be evaluated using Eq.(\ref{fluctuation}). We have used a constant field force here, directed solely towards the $x$ direction and denoted as $F_x$, with $F_y=0$. This results in the torque value as $\tau(t)=-y(t)F_x$. Putting this value of $\tau$ in the Eq.(\ref{taumsd}), we get
\begin{equation}
\begin{split}
    \langle\Delta\theta^2(t)\rangle&=\int_{0}^{t}dt^\prime\int_{0}^{t}dt^{\prime\prime}\langle\eta_3(t^\prime)\eta_3(t^{\prime\prime})\rangle+\Gamma^2_\theta F_x^2\int_{0}^{t}dt^\prime\int_{0}^{t}dt^{\prime\prime}\langle y(t^\prime)y(t^{\prime\prime})\rangle\\
    &=2D_\theta t+\Gamma^2_\theta F_x^2\langle y^2(t)\rangle
    \end{split}
    \label{finaltaumsd}
\end{equation}
Here $\langle y^2(t)\rangle$ is the MSD along $y$ axis of the particle under constant force field shown in the Eq.(\ref{MSDconsty}).
\end{appendix}
\end{document}